\documentclass[]{aa}
\usepackage{txfonts}
\usepackage{natbib}
\usepackage{amssymb}
\bibpunct{(}{)}{;}{a}{}{,}
\usepackage{epsf}
\usepackage{graphicx}
\usepackage{epsfig}
\usepackage{graphics}
\usepackage{subfigure}
\usepackage[english]{babel}
\usepackage{rotating}
\usepackage{color}

\newcommand{\teff}{$T_{\rm{eff}}$}
\newcommand{\logg}{$\log g$}
\newcommand{\lL}{\ifmmode \log \frac{L}{L_{\sun}} \else $\log \frac{L}{L_{\sun}}$\fi}

\newcommand{\vsini}{$V$~sin$i$}
\newcommand{\vinf}{$v_{\infty}$}

\newcommand{\vmac}{$v_{\rm mac}$}

\newcommand{\kms}{km~s$^{-1}$}
\newcommand{\msun}{M$_{\sun}$}

\begin{document}

\title{A study of the effect of rotational mixing on massive stars evolution: surface abundances of Galactic O7-8 giant stars }
\author{F. Martins\inst{1}
\and S. Sim\'on-D\'iaz\inst{2}\inst{3}
\and R.H. Barb\'a\inst{4}
\and R.C. Gamen\inst{5}
\and S. Ekstr$\ddot{\rm o}$m\inst{6}
}
\institute{LUPM, Universit\'e de Montpellier, CNRS, Place Eug\`ene Bataillon, F-34095 Montpellier, France  \\
\and
Instituto de Astrofísica de Canarias, 38200, La Laguna, Tenerife, Spain \\
\and
Departamento de Astrof\'isica, Universidad de La Laguna, E-38205 La Laguna, Tenerife, Spain \\
\and
Departamento de F\'{\i}sica y Astronom\'{\i}a, Universidad de La Serena, Av. Juan Cisternas 1200 N, La Serena, Chile\\
\and
Instituto de Astrof\'isica de La Plata, CONICET--UNLP, and Facultad de Ciencias Astron\'omicas y Geof\'isicas, UNLP. Paseo del Bosque s/n, La Plata, Argentina. \\
\and
Geneva Observatory, University of Geneva, 51 chemin des Maillettes, CH-1290 Sauverny, Switzerland
}

\offprints{Fabrice Martins\\ \email{fabrice.martins@umontpellier.fr}}

\date{Received / Accepted }

\abstract
{Massive star evolution remains only partly constrained. In particular, the exact role of rotation has been questioned by puzzling properties of OB stars in the Magellanic Clouds.}
{Our goal is to study the relation between surface chemical composition and rotational velocity, and to test predictions of evolutionary models including rotation.}
{We have performed a spectroscopic analysis of a sample of fifteen Galactic O7-8 giant stars. This sample is homogeneous in terms of mass, metallicity and evolutionary state. It is made of stars with a wide range of projected rotational velocities.}
{We show that the sample stars are located on the second half of the main sequence, in a relatively narrow mass range (25-40~\msun). Almost all stars with projected rotational velocities above 100 \kms\ have N/C ratios about ten times the initial value. Below 100 \kms\ a wide range of N/C values is observed. The relation between N/C and surface gravity is well reproduced by various sets of models. Some evolutionary models including rotation are also able to consistently explain slowly rotating, highly enriched stars. This is due to differential rotation which efficiently transports nucleosynthesis products and allows the surface to rotate slower than the core. In addition, angular momentum removal by winds amplifies surface braking on the main sequence. Comparison of the surface composition of O7-8 giant stars with a sample of B stars with initial masses about four times smaller reveal that chemical enrichment scales with initial mass, as expected from theory. }
{Although evolutionary models that  include rotation face difficulties in explaining the chemical properties of O- and B-type stars at low metallicity, some of them can consistently account for the properties of main-sequence Galactic O stars in the mass range 25-40 \msun.}


\authorrunning{Martins et al.}
\titlerunning{Surface abundances of O7-8~III stars}

\maketitle

\section{Introduction}
\label{s_intro}

The evolution of massive stars depends on several ingredients. As for all types of stars, initial mass is the main one. The relation between mass and luminosity implies that more massive stars are located higher in the Hertzsprung-Russell diagram. They live less long and develop stronger stellar winds. The associated mass-loss rates severely affect the life of massive stars, removing mass and leading to important changes in the internal structure and surface properties \citep{cm86}. Rotation plays also a key role on the evolution, changing not only the geometry of stars but also affecting the transport of angular momentum and chemical species \citep{mm00}. Finally, the presence of a companion can drastically modify the fate of massive stars because of tidal interactions and mass exchange \citep{langer12}. 

Comparisons between predictions of evolutionary models and results of spectroscopic analysis of observed stars is a standard way of testing and improving models, with the goal of a better knowledge of stellar evolution. Surface abundances are particularly interesting. They can be modified by stellar winds that remove external layers and reveal internal regions where nucleosynthesis products are present. Rotation can also modify the surface composition: mixing processes triggered by rotation can bring freshly processed material from the interior to the surface. Finally, mass accretion in binary systems obviously contaminates the surface properties of both components. 

Evolutionary models including rotation are being developed by several groups \citep{mema00,hlw00,brott11a,cl13}. In parallel, atmosphere models now routinely include metals, allowing determinations of C, N, and O (among others) abundances \citep{hil03,hunter08,morel08,ngc2244,rg12,jc13,mimesO}. Relatively large samples of stars are required to provide quantitative constraints on the predictions of evolutionary models. \citet{hunter08} presented surface nitrogen abundances for several tens of B stars in the Large Magellanic Cloud. Although the majority of stars showed a correlation between nitrogen content and projected rotational velocities, Hunter et al.\ also discovered that slowly rotating and N-rich and fast rotators and unevolved objects existed \citep[see also][]{morel06}. The former group questioned the role of rotational mixing in massive stars evolution since evolutionary models could not explain such strong chemical enrichment with so small projected rotational velocities. \citet{rg12} found similar slowly-rotating nitrogen-rich O stars in the Large Magellanic Cloud. \citet{maeder09} cautioned that surface composition does not depend only on rotation (see above) and stressed that isolating its effect required homogeneous samples. 

Analysis of O stars in different environments showed that evolutionary models usually reproduced their surface properties consistently \citep{ngc2244,jc12,jc13,mimesO}. In particular, \citet{mimesO} found that chemical enrichment was very well correlated with evolutionary state. A direct test of the relation between surface composition and projected rotational velocity was performed by \citet{rg12} for O stars in the Magellanic Clouds, but is still missing for Galactic O stars. In addition, samples analyzed so far cover wide ranges of mass and age that blur the effects of rotation.

The goal of the present study is to perform such an investigation. In Sect.\ \ref{s_obs} we describe how we built our sample and the related observations. Sect.\ \ref{s_mod} explains our method to determine the surface properties. The results are given in Sect.\ \ref{s_res} and discussed in Sect.\ \ref{s_disc}. Finally, we gather our conclusions in Sect.\ \ref{s_conc}.

\section{Sample and observations}
\label{s_obs}

Surface abundances of OB stars depend on several parameters: age (or evolutionary state), initial mass, metallicity, rotation rate, presence of a companion \citep{brott11a,ek12,cl13,maeder09}. To isolate the effects of rotation, it is thus necessary to reduce the space of the other parameters as much as possible. Consequently, we first excluded known close binaries from our sample since such objects are likely to interact and affect surface chemical composition \citep{langer12}. We then built our sample from Galactic stars bright enough to be observed at high spectral resolution on 1-4m telescopes. This ensures that the stars are relatively close to the Sun and share a common metallicity. 

As was shown by \citet{mimesO} surface nitrogen enrichment and carbon and oxygen depletion increase as stars move away from the zero-age main sequence and to the post-main sequence phase. This trend is also predicted by evolutionary models including rotation \citep[e.g.,][]{brott11a}. Surface abundances thus differ more from their initial values in supergiants than in dwarfs. However, supergiants are usually post-main sequence objects. Their radius has increased significantly at the end of the main sequence and consequently their rotational velocity has dropped. \citet{hunter08} showed that stars with surface gravities lower than 3.2 did not have \vsini\ larger than about 100 \kms, while for objects with higher surface gravities, \vsini\ was in the range $\sim$20 to 400 \kms\ \citep[see also][]{mp13}. Hence, in order to have a sample showing evidence of chemical mixing and covering a wide range of projected rotational velocities, giants appeared to be the best candidates. Assuming they are not binary merger products, these stars are already evolved from the ZAMS (and thus are chemically evolved) but have not yet reached the TAMS, ensuring that they can cover a range of \vsini. 

Finally, we wanted to build a sample with a narrow range of initial masses. In practice, we selected stars located in a relatively narrow region in the \logg\ -- \teff\ diagram (see below). This implied stars with similar effective temperatures and consequently similar spectral types. Given the shape of the initial mass function, early-type O stars are not as numerous as late-type O stars and we thus focussed on the latter.

The final selection was governed by the availability of high spectral resolution data in the IACOB and OWN surveys \citep{ss15,barba10}. At the time of our sample selection, O7-8 was the spectral type range with the largest number of data available. We thus decided to focus on them. We ended up with a sample of 15 objects with spectral type O7-8 and luminosity class II and III. The main characteristics of the spectroscopic data we used are given in Table.\ \ref{tab_obs}. For information on the data reduction, we refer the reader to the work of \citet{barba10,walborn11} and \citet{ss15}. 

The bulk of our sample is made of presumably single stars. In appendix \ref{ap_var} and Fig.\ \ref{fig_var} we show time series of spectra for most of the sample stars. BD~+60261, HD~171589 and HD~163800 are the only stars showing small radial velocity (RV) variations. All other targets are stable. Assuming that these three RV variable stars are binaries(such small RV variations may also be due to stellar oscillations) and that they have experienced mass transfer, we estimate that at most 20\% of our sample is contaminated by binarity effects. Regardless of RV variations, one can also quote \citet{demink14} who predict an incidence of 19\% of merger products among presumably single stars.

Our sample remains small but is built to limit the effects of initial mass, age, binarity and metallicity on surface abundances. It covers a wide range of projected rotational velocities (see Section \ref{s_mod}) and allows to test the effects of rotational mixing on Galactic O stars.

\begin{table*}
\begin{center}
\caption{Observational information.} \label{tab_obs}
\begin{tabular}{lcccc}
\hline
Star        & Sp.T.$^1$    &  Instrument & Resolving power & Date of observation\\    
            &           & \\
\hline
BD~+60261    & O7.5 III(n)(f)   & FIES    & 25000 & 19 sep 2011 \\
HD~24912     & O7.5 III(n)(f)   & FIES    & 46000 & 09 sep 2008 \\
HD~34656     & O7.5 II(f)       & FIES    & 46000 & 24 dec 2012 \\
HD~35633     & O7.5 II(n)(f)$^2$& FIES    & 25000 & 29 jan 2013 \\
HD~36861     & O8 III((f))      & HERMES  & 85000 & 03 feb 2014 \\
HD~94963     & O7 II(f)         & FEROS   & 48000 & average 21 \& 27 apr 2007 \\
HD~97434     & O7.5III(n)((f))  & FEROS   & 48000 & average 26 may 2007 / 22 mar 2011 / 02 apr 2015 \\
HD~151515    & O7 II(f)         & FEROS   & 48000 & 19 apr 2007 \\
HD~162978    & O8 II((f))       & FIES    & 46000 & 28 aug 2011 \\
HD~163800    & O7.5 III((f))    & FIES    & 46000 & 10 sep 2011 \\
HD~167659    & O7 II-III(f)     & FIES    & 46000 & 08 sep 2010 \\
HD~171589    & O7 II(f)         & FIES    & 25000 & 12 sep 2011 \\
HD~175754    & O8II(n)((f))p    & FIES    & 46000 & 08 sep 2010 \\
HD~186980    & O7.5 III(f)      & FIES    & 46000 & 08 aug 2010 \\
HD~203064    & O7.5 IIIn((f))   & HERMES  & 85000 & 16 jun 2011 \\
\hline
\end{tabular}
\tablefoot{1: Spectral types are from \citet{sota11,sota14}. 2: from J. Ma\'iz Apell\'aniz, private communitation.}
\end{center}
\end{table*}

\section{Modeling and spectroscopic analysis}
\label{s_mod}

We used the code CMFGEN to perform the spectroscopic analysis. CMFGEN compute non-LTE atmosphere models in a spherical geometry. It includes stellar winds and line-blanketing \citep{hm98}. The radiative transfer, rate equations and energy conservation equation are solved iteratively to provide the temperature structure and the level populations. The following elements are taken into account in the calculations: H, He, C, N, O, Ne, Mg, Si, S, Ar, Ca, Fe, Ni. The radiative force is computed self-consistently and is used to perform iterations of the hydrodynamical equation below the sonic point, in order to obtain a consistent hydrodynamical structure in the inner atmosphere. The resulting velocity law is connected to a $\beta$ law in the outer part: $v = v_{\infty} (1-\frac{R}{r})^{\beta}$ with $R$ the stellar radius and \vinf\ the terminal velocity (we adopted $\beta$ = 1.0).
Once the atmosphere model is converged, we compute the corresponding synthetic spectrum from a formal solution of the radiative transfer equation. In that step, a microturbulence varying from 10 \kms\ at the photosphere to $0.1 \times\ v_{\infty}$ at the outer boundary is adopted. \\
The spectroscopic analysis based on such synthetic spectra is performed as follows.

\subsection{Line-broadening parameters (\vsini\ and \vmac)}

These two parameters were obtained simultaneously using the {\sc iacob-broad} tool and following the strategy described in \citet{sergio14}. We used \ion{O}{iii}\,5592 as diagnostic line for all the considered targets. From the various options provided by {\sc iacob-broad}, we kept the values \vsini\ and \vmac\ determined from the goodness of fit solution (after checking the good agreement between \vsini(FT) and \vsini(GOF)), as published by \citet{sergio14}.

\subsection{Effective temperature and surface gravity}

\teff\ is constrained from the ionization balance method, (i.e., the relative strength of \ion{He}{i} and \ion{He}{ii} lines). We used \ion{He}{i}~4026 (blend with \ion{He}{ii}~4026), \ion{He}{i}~4388, \ion{He}{i}~4471, \ion{He}{i}~4713, \ion{He}{i}~4921, \ion{He}{i}~5015, \ion{He}{ii}~4200, \ion{He}{ii}~4542, and \ion{He}{ii}~5412. Surface gravity was obtained from the width of Balmer lines:  H$\beta$, H$\gamma$, H$\delta$, H$\epsilon$, H$\zeta$, and H$\eta$. \teff\ and \logg\ were obtained simultaneously: a grid of synthetic spectra was convolved with a rotational and radial-tangential profile (parameterized by \vsini\ and \vmac). Each spectra was subsequently compared to the observed one and the goodness of fit was quantified by a $\chi^2$ calculation. The best fit model was extracted from this process. The error bars in \teff\ and \logg\ are correlated, but on average \teff\ is estimated at better than 2000 K and \logg\ at better than 0.2 dex.  

\subsection{Surface abundances}

For each star, a set of models with \teff\ and \logg\ determined as explained above but with different compositions in C, N and O were calculated. Comparison to observed CNO lines was made (quantified by a $\chi^2$ analysis) to provide the final abundances. We used mainly \ion{C}{iii}~4070 for carbon, complemented when possible by \ion{C}{iii}~4153, \ion{C}{iii}~4157  and \ion{C}{iii}~4163. For nitrogen, \ion{N}{iii}~4511, \ion{N}{iii}~4515, \ion{N}{iii}~4518, and \ion{N}{iii}~4524 were the main diagnostics. \ion{N}{iii}~4196 was used in a few cases. Finally, the oxygen abundance was obtained from \ion{O}{iii}~3962 and \ion{O}{iii}~5592, with the occasional use of \ion{O}{iii}~3791. Due to the small number of lines used, we can only provide a lower limit on O/H (above this value, $\chi^2$ reaches a minimum plateau).

\vspace{0.3cm}

For two objects (HD~34656 and HD~175754) we had to increase the helium content from 0.10 to 0.15 to correctly reproduce the helium lines. For all the other targets, He/H=0.10 was adopted. In absence of accurate distance estimates for the targets, we adopted their luminosity from the calibrations of \citet{msh05}.

\section{Results}
\label{s_res}

The results of the quantitative analysis are summarized in Table \ref{tab_param}. The best fits for each star are shown in Appendix \ref{ap_fit}. Six stars (HD~24912, HD~34656, HD~36861, HD~162978, HD~186980, HD~203064) were previsously studied by \citet{mimesO}. The parameters we determined in the present study are in good agreement with those of \citet{mimesO}. Within the error bars, the surface abundances are the same between both study, with the only exception of C/H for HD~203064. Given the different set of data and different assumptions regarding \vsini\ and \vmac\ this indicates that our results are safe.

\begin{table*}
\begin{center}
\caption{Parameters of the sample stars.} \label{tab_param}
\begin{tabular}{lcccccccrrrrr}
\hline
Star        & Sp.T.   & Teff &   logg & logg$_c$&  \vsini\  &  \vmac  &   C/H         &    N/H      & O/H                    &  $\log$(N/C)   &  $\log$(N/O)  \\    
            &            & [kK] &        &         &  [\kms]   &  [\kms] & [10$^{-4}$]  &  [10$^{-4}$] &  [10$^{-4}$]             &                &                \\
\hline
BD~+60261    & O7.5 III(n)(f) & 34.5  & 3.50  & 3.54 & 162   & $<$110 & 2.3$^{+0.5}_{-0.5}$ & 5.8$^{+3.1}_{-3.1}$ & $>$4.0           & 0.40$^{+0.25}_{-0.25}$           & $<$0.16 \\ 
HD~24912     & O7.5 III(n)(f) & 33.5  & 3.50  & 3.56 & 230   & $<$80  & 1.3$^{+0.2}_{-0.2}$ & 5.1$^{+1.0}_{-1.0}$ & 2.3$^{+1.3}_{-0.5}$ & 0.59$^{+0.11}_{-0.11}$           & 0.35$^{+0.26}_{-0.13}$ \\
HD~34656     & O7.5 II(f)     & 35.5  & 3.60  & 3.61 & 63    & 73     & 0.9$^{+0.3}_{-0.3}$ & 7.8$^{+4.7}_{-3.2}$ & 3.6$^{+1.2}_{-1.2}$ & 0.94$^{+0.30}_{-0.23}$           & 0.34$^{+0.30}_{-0.23}$\\ 
HD~35633     & O7.5 II(n)(f)  & 33.5  & 3.40  & 3.45 & 168   & $<$125 & 2.1$^{+0.5}_{-0.4}$ & 1.0$^{+0.6}_{-0.6}$ & $>$4.7           & -0.32$^{+0.28}_{-0.27}$           & $<$-0.67 \\ 
HD~36861     & O8 III((f))    & 35.0  & 3.75  & 3.75 & 60    & 60     & 2.3$^{+0.6}_{-0.6}$ & 1.4$^{+0.6}_{-0.5}$ & $>$3.5           & -0.22$^{+0.22}_{-0.19}$           & $<$-0.40\\ 
HD~94963     & O7 II(f)       & 35.0  & 3.50  & 3.51 & 63    & 99     & 2.0$^{+0.6}_{-0.4}$ & 4.7$^{+1.7}_{-1.7}$ & $>$3.2           & 0.37$^{+0.20}_{-0.18}$           & $<$0.17\\
HD~97434     & O7.5III(n)((f))& 34.0  & 3.50  & 3.57 & 230   & $<$125 & 1.3$^{+0.2}_{-0.3}$ & 5.5$^{+5.0}_{-2.2}$ & 9.2$^{+6.1}_{-3.5}$ & 0.63$^{+0.40}_{-0.20}$           & -0.22$^{+0.49}_{-0.24}$\\ 
HD~151515    & O7 II(f)       & 36.0  & 3.60  & 3.61 & 67    & 88     & 1.7$^{+0.6}_{-0.6}$ & 5.6$^{+4.0}_{-2.8}$ & 2.5$^{+1.7}_{-4.7}$ & 0.52$^{+0.35}_{-0.27}$           & 0.35$^{+0.87}_{-0.37}$\\
HD~162978    & O8 II((f))     & 33.5  & 3.40  & 3.41 & 53    & 93     & 1.7$^{+0.2}_{-0.2}$ & 5.9$^{+3.3}_{-2.3}$ & 4.0$^{+5.5}_{-2.3}$ & 0.54$^{+0.25}_{-0.18}$           & 0.17$^{+0.64}_{-0.30}$\\ 
HD~163800    & O7.5 III((f))  & 35.5  & 3.60  & 3.61 & 56    & 100    & 3.0$^{+0.4}_{-0.4}$ & 3.1$^{+0.8}_{-0.8}$ & $>$6.4           & 0.01$^{+0.13}_{-0.13}$           & $<$-0.31\\ 
HD~167659    & O7 II-III(f)   & 36.0  & 3.60  & 3.61 & 76    & 92     & 2.2$^{+0.3}_{-0.3}$ & 2.1$^{+0.6}_{-0.6}$ & 4.5$^{+3.5}_{-1.8}$ & -0.02$^{+0.14}_{-0.12}$           & -0.33$^{+0.36}_{-0.20}$\\ 
HD~171589    & O7 II(f)       & 36.0  & 3.75  & 3.76 & 98    & 83     & 2.0$^{+0.6}_{-0.8}$ & 5.5$^{+2.0}_{-2.0}$ & 5.4$^{+4.0}_{-1.7}$ & 0.44$^{+0.20}_{-0.23}$           & 0.01$^{+0.36}_{-0.21}$\\
HD~175754    & O8II(n)((f))p  & 33.0  & 3.30  & 3.37 & 191   & $<$90  & 2.6$^{+0.8}_{-0.5}$ & 9.6$^{+11.0}_{-5.9}$ & 7.3$^{+4.2}_{-2.7}$& 0.57$^{+0.52}_{-0.28}$           & 0.12$^{+0.56}_{-0.31}$ \\
HD~186980    & O7.5 III(f)    & 35.5  & 3.75  & 3.75 & 52    & 96     & 2.0$^{+0.3}_{-0.3}$ & 2.8$^{+0.8}_{-0.8}$ & 6.0$^{+3.3}_{-1.7}$ & 0.15$^{+0.14}_{-0.14}$           & -0.33$^{+0.27}_{-0.17}$\\
HD~203064    & O7.5 IIIn((f)) & 33.5  & 3.60  & 3.68 & 314   & $<$45  & 0.7$^{+0.4}_{-0.4}$ & 1.8$^{+0.6}_{-1.0}$ & $>$5.5           & 0.41$^{+0.35}_{-0.29}$           & $<$-0.49 \\ 
\hline
\end{tabular}
\tablefoot{Uncertainties on \teff, \logg, \vsini\ and \vmac\ are $\sim$2.0kK, 0.20 dex, 10 and 20 \kms\ respectively. logg$_c$ is the surface gravity corrected for centrifugal acceleration. Abundances are number ratios.}
\end{center}
\end{table*}

\begin{figure}[t]
\centering
\includegraphics[width=9cm]{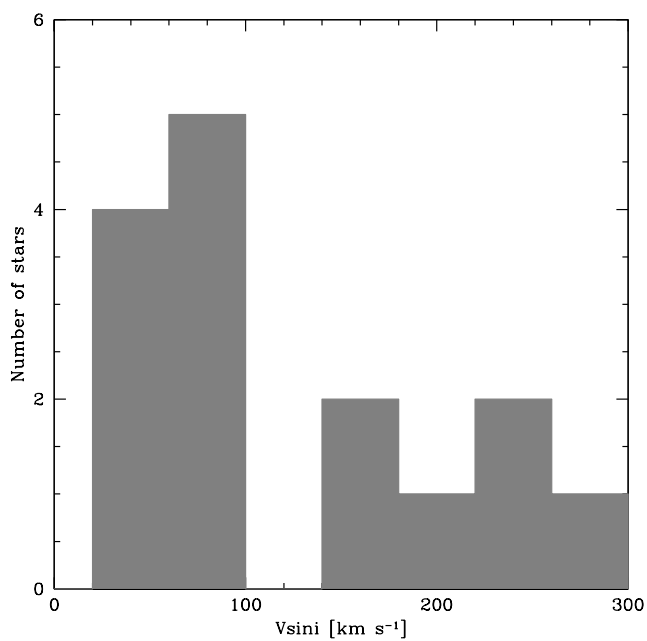}
\caption{Distribution of projected rotational velocities (\vsini) in our sample.}
\label{v_dist}
\end{figure}

Figure\ \ref{v_dist} gives the distribution of \vsini\ in our sample. A wide range of values is covered, from about 50 to nearly 300 \kms. Nine (six) stars have projected rotational velocities smaller (larger) than 100 \kms. Obviously the distribution of true rotational velocities may be different owing to the projection factor. However, it is unlikely that all stars with low \vsini\ are fast rotators seen nearly pole-on.

\begin{figure}[t]
\centering
\includegraphics[width=9cm]{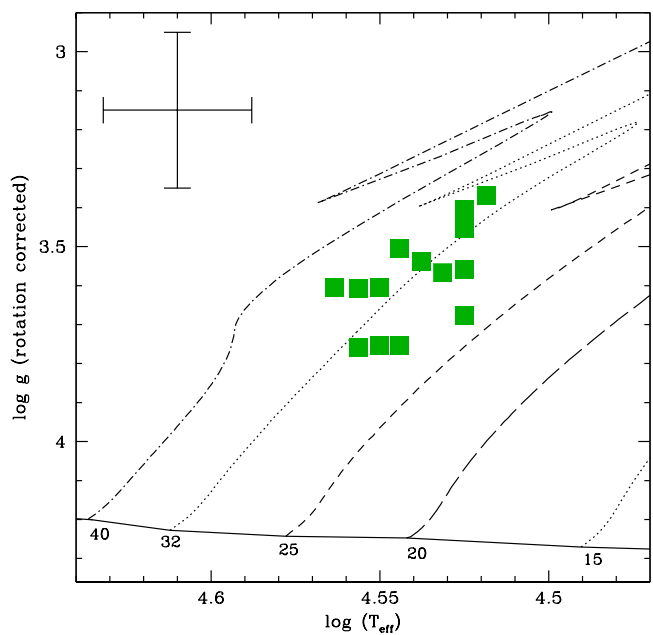}
\caption{\logg\ - log(\teff) diagram for the sample stars. Typical uncertainties are shown in the upper-left corner. Evolutionary tracks from \citet{ek12} including rotation are overplotted, labeled by initial masses.}
\label{fig_hr}
\end{figure}

The position of the target stars in the \logg\ - \teff\ diagram is shown in Fig.\ \ref{fig_hr}. They are located over a relatively small area, as expected from the selection criteria: their \teff\ and \logg\ do not differ by more than 3000 K and 0.40 dex respectively. According to the evolutionary tracks of \citet{ek12}, all targets have initial masses between 25 and 40 \msun. They are located in the second part of the main sequence, but have not yet reached the terminal-age main sequence (TAMS). All stars are thus in a relatively close state of evolution.

\begin{figure}[t]
\centering
\includegraphics[width=9cm]{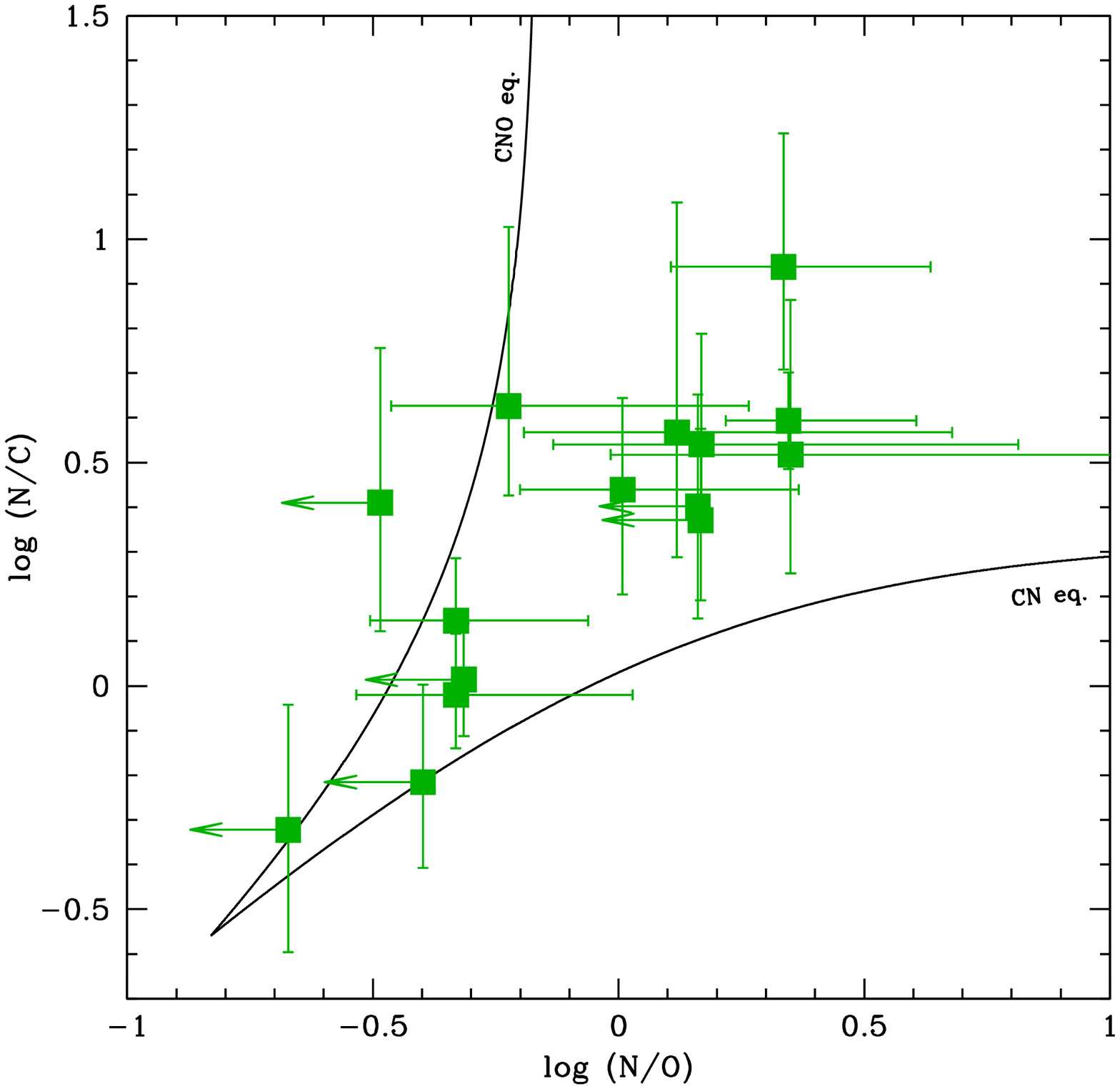}
\caption{log(N/C) - log(N/O) diagram for the sample stars. Solid lines indicate the prediction of nucleosynthesis through the partial CN or complete CNO cycle.}
\label{nco}
\end{figure}

Fig.\ \ref{nco} shows the ratio N/C as a function of N/O. These two ratios are clearly correlated and lie between the predictions of complete CNO or partial CN cycles, as expected from nucleosynthesis predictions \citep[e.g.,][]{maeder14}. This indicates that the surface CNO abundances correspond to material produced in the stellar core and brought to the surface by mixing processes. Some stars are barely chemically evolved while others show nitrogen enrichment and carbon--oxygen depletion.

\begin{figure}[t]
\centering
\includegraphics[width=9cm]{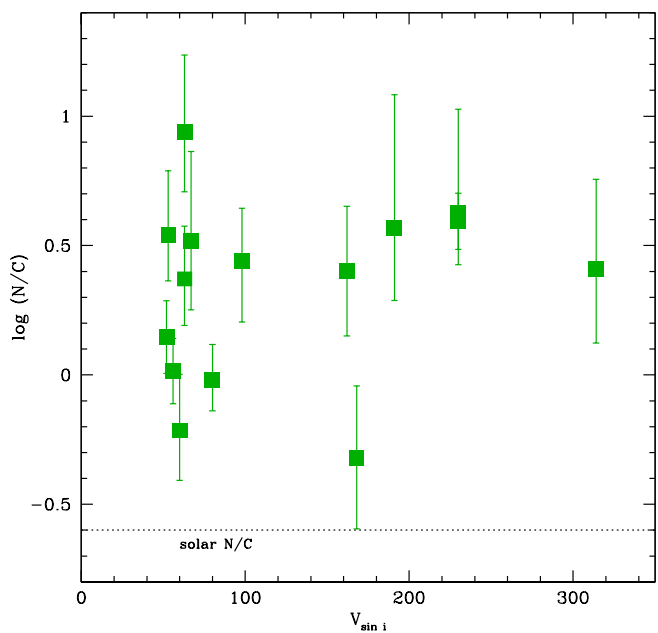}
\caption{log(N/C) - \vsini\ diagram for the sample stars. The dot-dashed horizontal line shows the solar N/C ratio. The lack of stars with \vsini\ $<$ 50 \kms\ is partly due to limitations of the method to determine \vsini\ \citep[see][]{sergio14}}
\label{nc_vsini_obs}
\end{figure}

\begin{figure}[t]
\centering
\includegraphics[width=9cm]{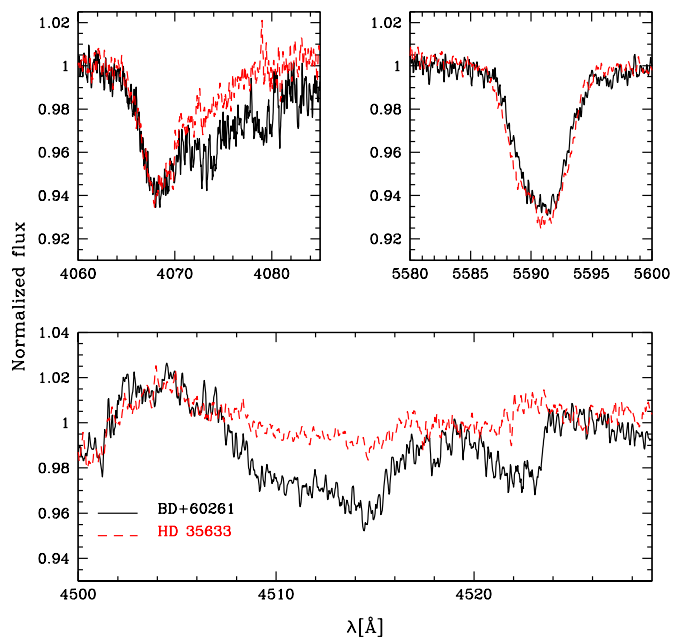}
\caption{Comparison between the observed spectrum of BD~+60261 (black solid line) and HD~35633 (red dashed line). \ion{C}{iii}~4070, \ion{O}{iii}~5592 and the \ion{N}{iii} lines around 4510 \AA\ are displayed in the upper left, upper right and bottom panel respectively. Both stars have very similar \vsini.}
\label{comp_v180}
\end{figure}

We have checked that our sample is made of stars with relatively homogeneous masses and evolutionary states. The targets also show various degrees of chemical enrichment, that can now be confronted to rotational velocities. This is shown in Fig.\ \ref{nc_vsini_obs}. The large majority of stars are N-rich and C-poor, but the range of N/C covers more than one order of magnitudes. There is no clear correlation between N/C and \vsini. Above 100 \kms, most stars have $\log (N/C) = 0.5$. The only exception is HD~35633 which shows very little enrichment. Figure\ \ref{comp_v180} shows a qualitative view of this, by comparing CNO lines of BD~+26261 and HD~35633. Both stars have similar \ion{C}{iii}~4070 and \ion{O}{iii}~5592 indicating the same C/H and O/H ratios (see Table \ref{tab_param}). But the \ion{N}{iii} lines are stronger in BD~+60261, leading to a larger N/C ratio. 
Below 100 \kms\ a wide range of N/C ratios is obtained for a relatively narrow range of projected rotational velocities (50-100 \kms).

\section{Discussion}
\label{s_disc}

\subsection{Testing evolutionary models for Galactic O stars}
\label{s_disc_ev}

The determination of CNO surface abundances in our sample of O7-8 giant stars reveals trends with projected rotational velocity already observed for lower mass B stars. \citet{hunter08} studied the nitrogen abundance of B-type stars in the Large Magellanic Cloud and concluded that among main sequence stars, a variety of N/H ratios were obtained. The majority of stars showed a trend of higher N content with larger rotational velocity. But some fast rotating objects were unevolved whereas some slow rotators were very N-rich. Both groups of objects questioned the ability of rotational mixing in accounting for surface abundances, since evolutionary models including rotation predict a correlation of N/H with \vsini. Subsequent analysis of B stars in the Small Magellanic Cloud also revealed the presence of N-rich stars with low \vsini\ \citep{hunter09}. The Galactic sample of B stars of \citet{hunter09} did not show such objects. 

As discussed by \citet{maeder09} \citep[see also Fig.~8 of][]{brott11a}, the dependence of surface abundances on multiple parameters (initial mass, age, metallicity, rotation rate) may be hampering the correct interpretation of the observed distribution of stars in the N/H -- \vsini\ diagram obtained by these studies. Taking this into account, \citet{brott11b} built population synthesis models and confirmed the results of \citet{hunter08}: a non negligible fraction of stars show either large chemical enrichment and small projected rotational velocities, or almost no enrichment and fast rotation. These objects could not be fully reproduced by the evolutionary models of \citet{brott11a}.

\begin{figure}[t]
\centering
\includegraphics[width=9cm]{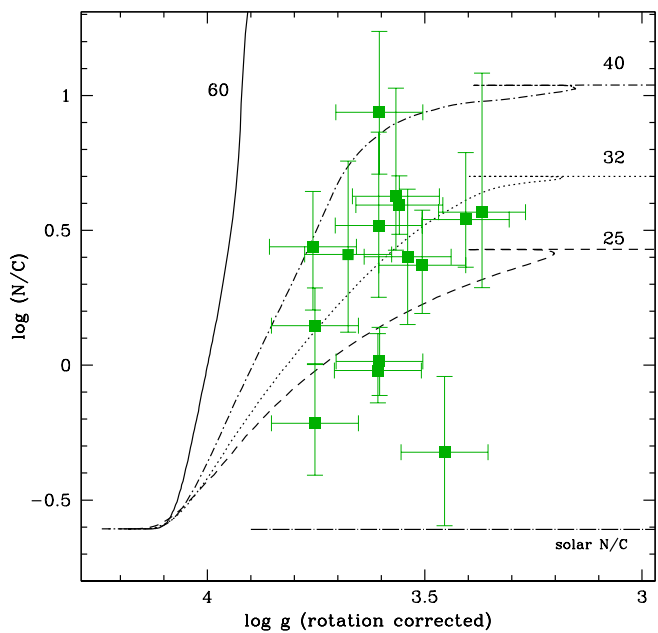}
\caption{$\log (N/C)$ as a function of \logg\ for the sample stars. The evolutionary tracks including rotation of \citet{ek12} are overplotted and labelled by initial mass (initial rotation $\sim$300 \kms). The horizontal dot-dashed line indicates the solar N/C ratio.}
\label{NC_logg}
\end{figure}

To see if such results were still valid in the mass range covered by O stars, \citet{jc13} confronted the surface abundances of SMC O stars to the predictions of the evolutionary models of \citet{brott11a} in a N/C versus \lL\ diagram. They concluded that in this diagram evolutionary models were able to account for the properties of the sample stars. At the same time, they found that in the N/H -- \vsini\ diagram, there was a group of N-rich slow rotators, as in the B stars studied by \citet{hunter08,hunter09} and the O stars analyzed by \citet{rg12}. 

More recently \citet{mimesO} analyzed the surface properties of seventy four Galactic O stars. They established a clear trend of stronger chemical enrichment as stars evolve from dwarfs to giants and supergiants. Evolutionary tracks including rotation produced by \citet{ek12} and \citet{cl13} could reproduce more than 80\% of the surface abundances in a $\log (N/C) - \log g$ diagram. The models of \citet{brott11a} reach high N/C too early (i.e., at too high \logg) to correctly reproduce the observed abundances. A trend of stronger enrichment in more massive stars was also highlighted by \citet{mimesO}. These results indicates that some of the current generation of evolutionary models including rotation are able to account quantitatively for the relation between surface abundances and evolutionary state. 

Figure\ \ref{NC_logg} confirms these findings for the present study: almost all of the O7-8 giant stars lie between the 25 and 40 \msun\ tracks of \citet{ek12}, in perfect agreement with the estimated initial mass of these objects from Fig.\ \ref{fig_hr}. The Geneva evolutionary tracks including rotation are thus able to reproduce the observed range of surface abundances in a $\log N/C - \log g$ diagram. The only clear outlier is HD~35633 that we discussed in Sect.\ \ref{s_res}.

\begin{figure*}[t]
\centering
\includegraphics[width=9cm]{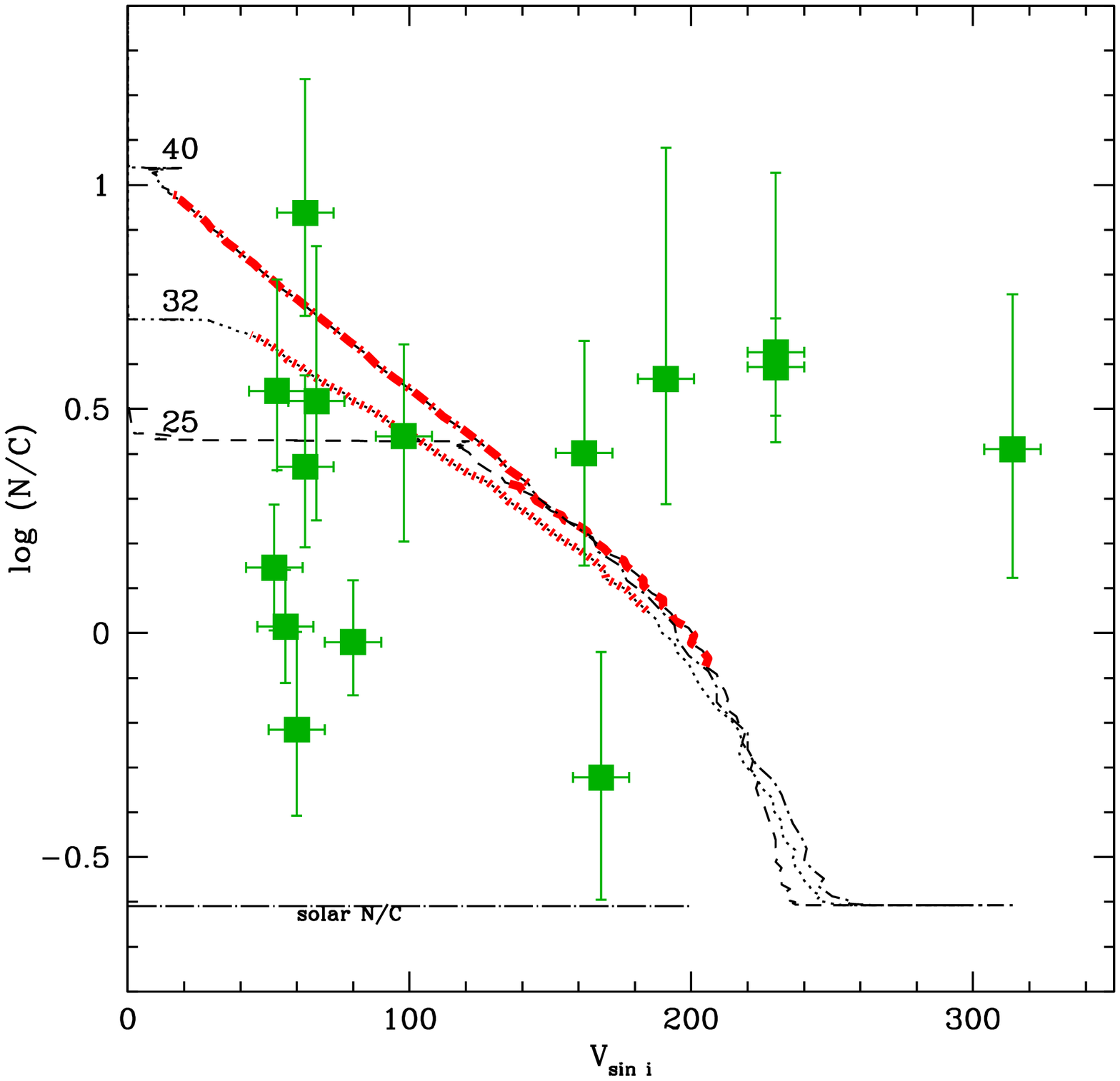}
\includegraphics[width=9cm]{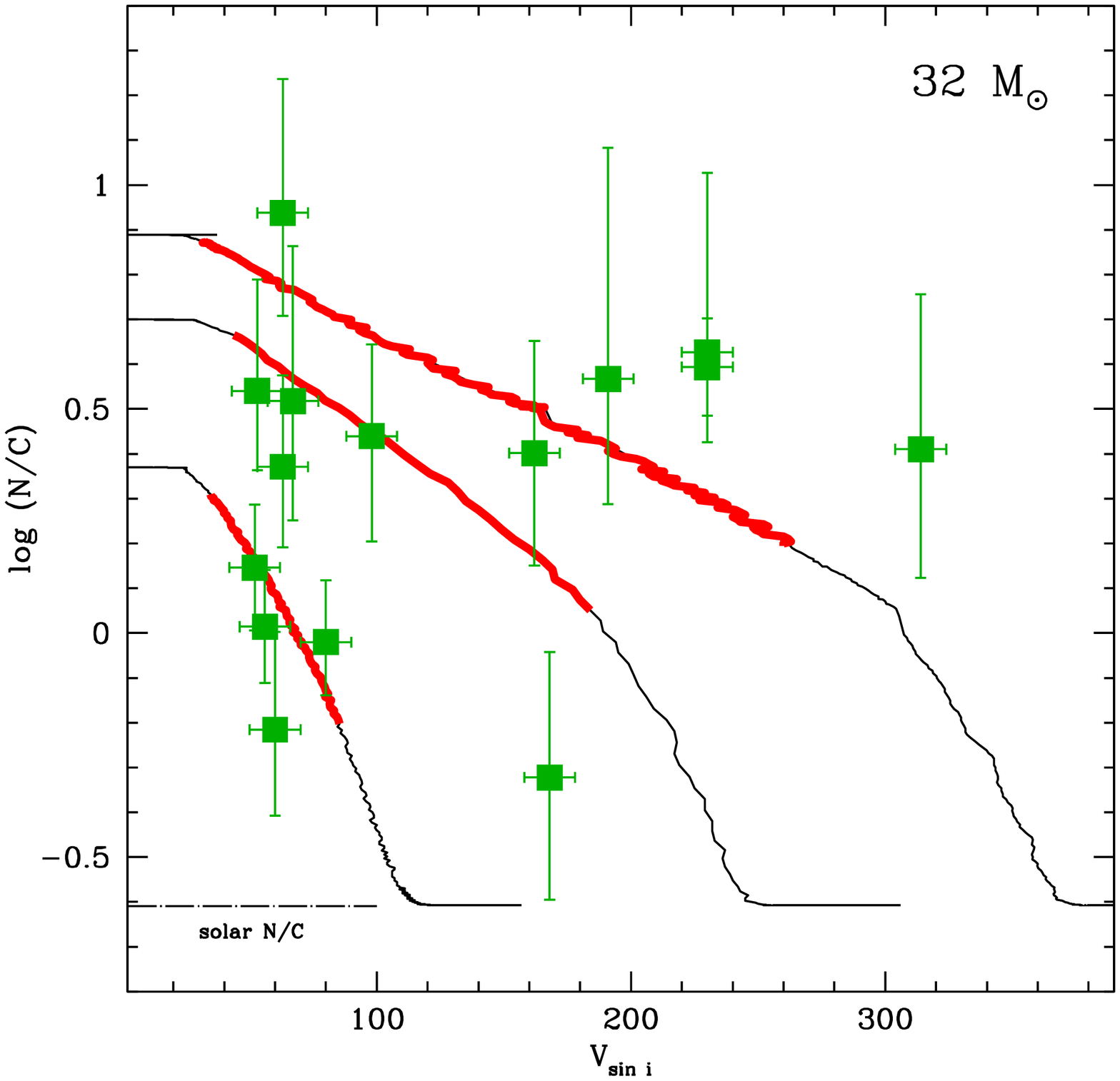}
\caption{As Fig.\ \ref{nc_vsini_obs} with evolutionary models including rotation from \citet{ek12}. The red bold part of the tracks corresponds to 3.3 < \logg\ < 3.7, the range covered by the sample stars. \textit{Left panel}: evolutionary tracks with initial masses 25, 32 and 40 \msun\ and an initial rotational velocity of $\sim$ 300 \kms. \textit{Right panel}: evolutionary tracks with an initial mass of 32 \msun\ and various initial rotational velocities.}
\label{nc_vsini}
\end{figure*}

The studies of \citet{hunter08,hunter09} and \citet{brott11b} showed that the evolutionary models of \citet{brott11a} could not reproduce the position of main sequence stars in the log(N/H)--\vsini\ diagram and for B stars in the Magellanic Clouds. Fig.\ \ref{nc_vsini} shows a similar diagram (log(N/C)--\vsini) for our sample stars, together with the predictions of \citet{ek12}. The velocities from the models are equatorial velocities and are thus upper limits on \vsini. The positions of the sample stars should be compared with the red bold part of the tracks which correspond to their range of surface gravities as deduced from Fig.\ \ref{fig_hr}. In the left panel of Fig.\ \ref{nc_vsini} all evolutionary models have an initial rotational velocity on the zero-age main sequence (ZAMS) close to 300~\kms. These models cover the range of observed N/C. Of particular interest is the ability of the models to reproduce slowly-rotating chemically enriched O stars. Indeed, braking during the main sequence is sufficiently strong to significantly reduce the equatorial velocity. At the same time, chemical enrichment is strong enough to produce large values of N/C. The red bold part of the 32 and 40 \msun\ evolutionary tracks crosses the region of highly enriched, slowly rotating stars. On the other hand the 25 \msun\ track remains with equatorial velocities larger than 100 \kms during the entire main sequence. With an average reduction of the equatorial velocity by a factor $4/\pi$ due to the $sin i$ factor, it can nonetheless reach values consistent with observations. Consequently tracks selected to represent the position of the sample stars in the \logg\ -- \teff\ diagram are also (partly) able to reproduce their positions in the upper left part of the log(N/C) -- \vsini\ diagram.

Fast rotating, chemically enriched stars, as well as slow rotators with little surface enrichment, can be accounted for by models with higher (respectively lower) initial rotational velocity. This is illustrated in the right panel of Fig.\ \ref{nc_vsini}. The models all have an initial mass of 32 \msun\ but different initial rotational velocities: in addition to an initial ratio of equatorial velocity to critical velocity of 0.4, we have computed models with ratios of 0.2 and 0.6 \citep[using the Geneva evolutionary code described in][]{ek12}. The part of the tracks matching the surface gravity of the sample stars nicely reproduces the range of N/C values. Hence, our study shows that in the mass range probed, the evolutionary models of \citet{ek12} are able to reproduce consistently the surface chemical composition of Galactic O stars, their position in the \logg\ -- \teff\ diagram and their projected rotational velocity.

This result is different from what was obtained for O and B stars in the Magellanic Clouds. \citet{hunter08,hunter09}, \citet{brott11b}, and \citet{rg12} used the evolutionary tracks of \citet{brott11a}. These models are based on the diffusive formalism of \citet{es78} to include rotational mixing and transport. In addition, they include magnetic fields which tend to lead to solid-body rotation. Angular momentum removal by stellar winds takes place, but due to the strong coupling between the stellar core and surface, angular momentum is extracted from the core and efficiently transported to the surface. This compensates for the angular momentum loss by winds. Hence, surface velocity remains high on the main sequence. \citet{ek12} use the advecto-diffusive formalism of \citet{zahn92} and \citet{mz98} for rotation and do not include magnetic fields. This produces models with differential rotation in which the coupling between core and surface is not as strong as in the case of solid-body rotation. Consequently surface velocity can decrease significantly already on the main sequence. The studies of \citet{hunter08,hunter09}, \citet{brott11b} and \citet{rg12} are also for low metallicity stars in which winds are weaker and rotational mixing stronger, while in the present study we analyze Galactic stars.

\begin{figure*}[t]
\centering
\includegraphics[width=0.47\textwidth]{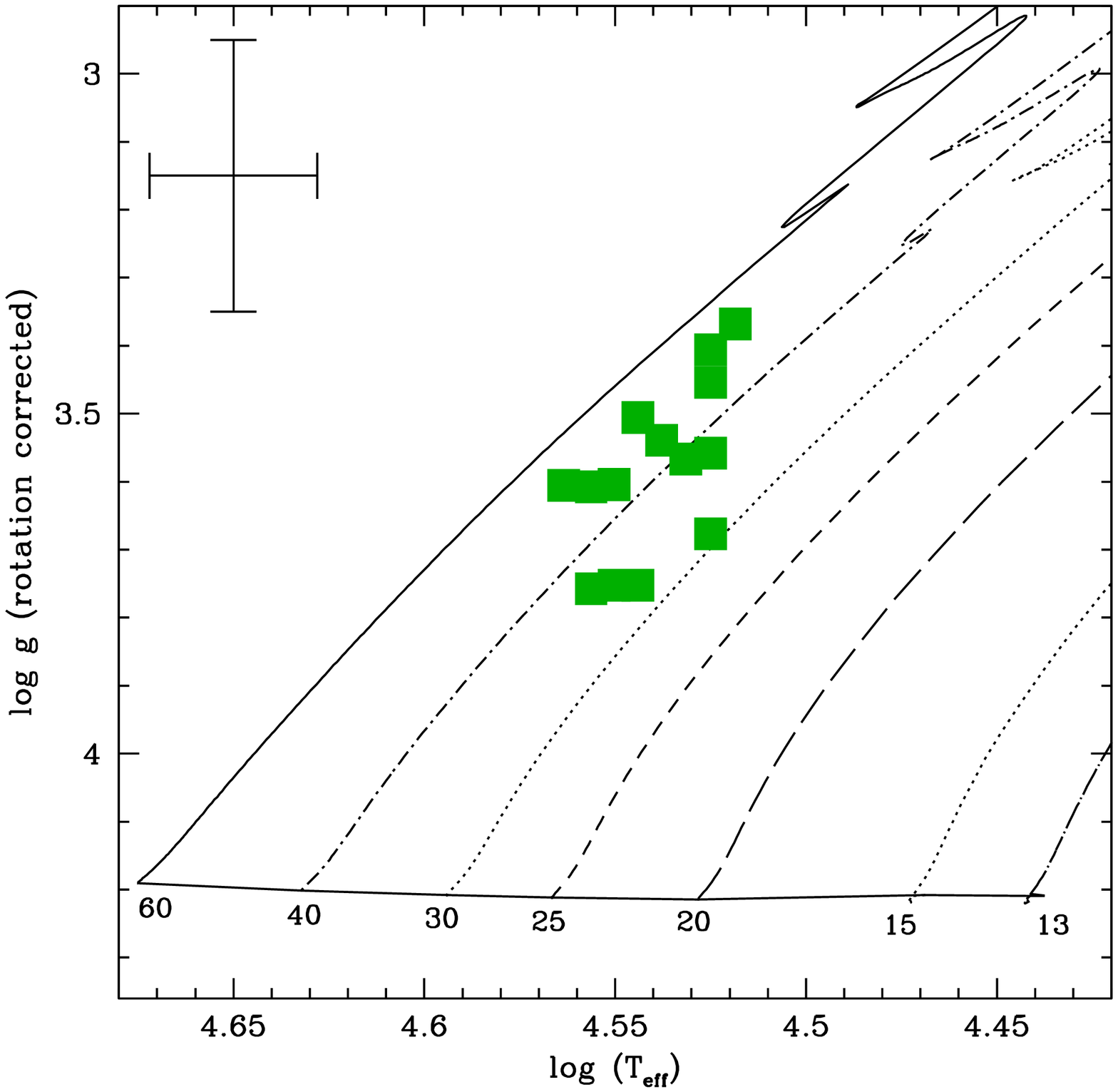}
\includegraphics[width=0.47\textwidth]{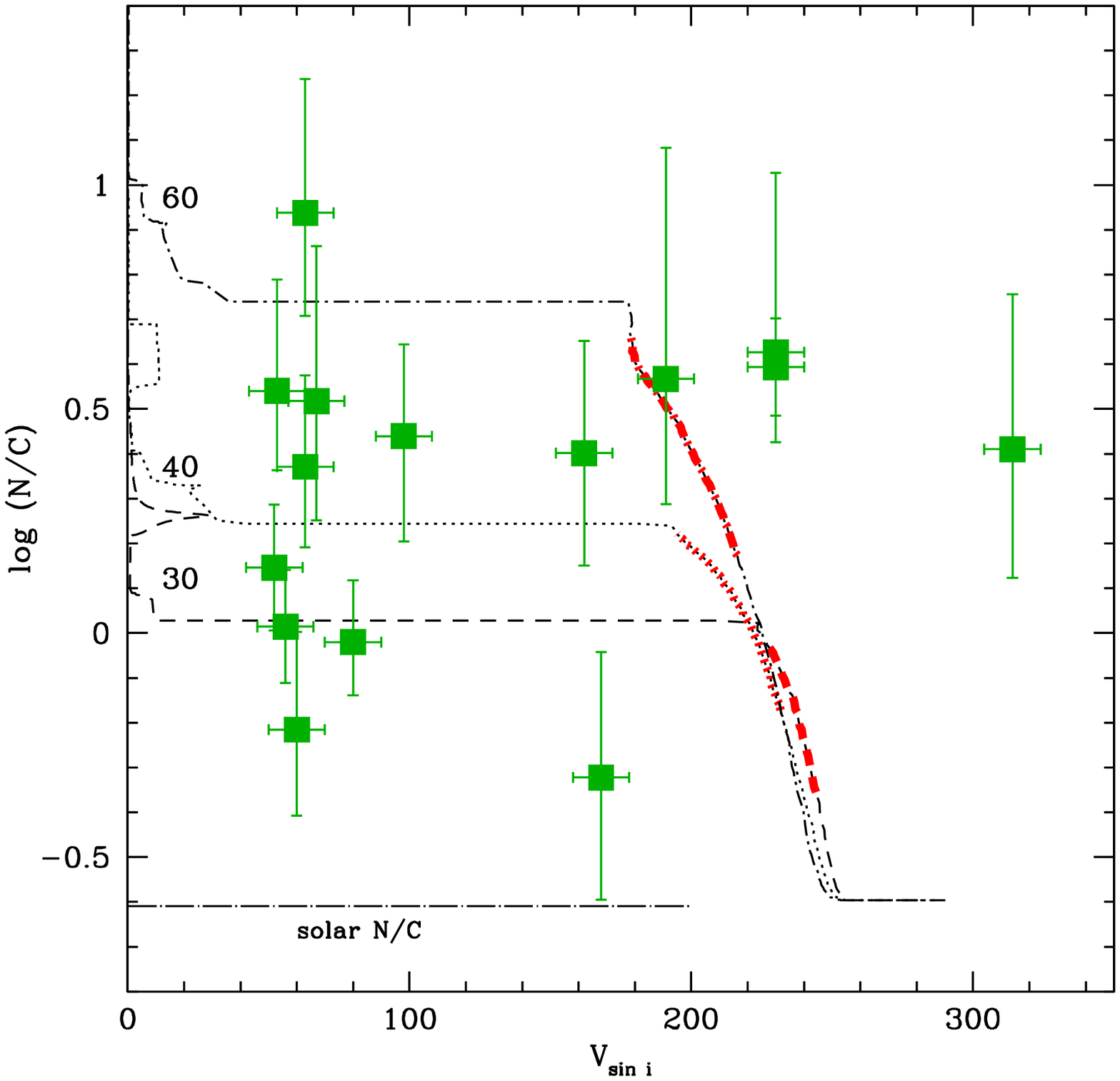}
\caption{As Fig.\ \ref{fig_hr} (left panel) and Fig.\ \ref{nc_vsini} (right panel) with the evolutionary models of \citet{cl13}.}
\label{comp_cl13}
\end{figure*}

A further test of evolutionary models is shown in Fig.\ \ref{comp_cl13}. The tracks of \citet{cl13} have now replaced those of \citet{ek12}. According to these new models, our sample stars have an initial mass between 30 and 60 \msun\ (see left panel of Fig.\ \ref{comp_cl13}). In the log(N/C) -- \vsini\ diagram the 30, 40 and 60 \msun\ tracks cover the range of observed N/C ratios. However, the part of the tracks corresponding to the surface gravities derived from the \logg\ -- \teff\ diagram remain at too high velocities to account for the low \vsini/high N/C stars, very much like the situation encountered for OB stars at low metallicity. We conclude that the models by \citet{cl13} are not braked sufficiently to provide a consistent picture of surface abundances in main sequence stars. The main difference between these models and those of \citet{ek12} is the prescription for the shear mixing coefficient: \citet{maeder97} for the latter, \citet{tz97} for \citet{cl13}. Since otherwise the inclusion of rotational mixing is the same and mass loss rates are taken from the same source \citep{vink01}, we attribute the difference in angular momentum transport to this different shear mixing coefficient.

The main conclusion of our study is that some (but not all) evolutionary models are able to consistently reproduce the surface properties and abundances of Galactic O stars in the range 25-40 \msun.

\subsection{Comparison with Galactic B stars}
\label{s_B}

\begin{figure}[t]
\centering
\includegraphics[width=9cm]{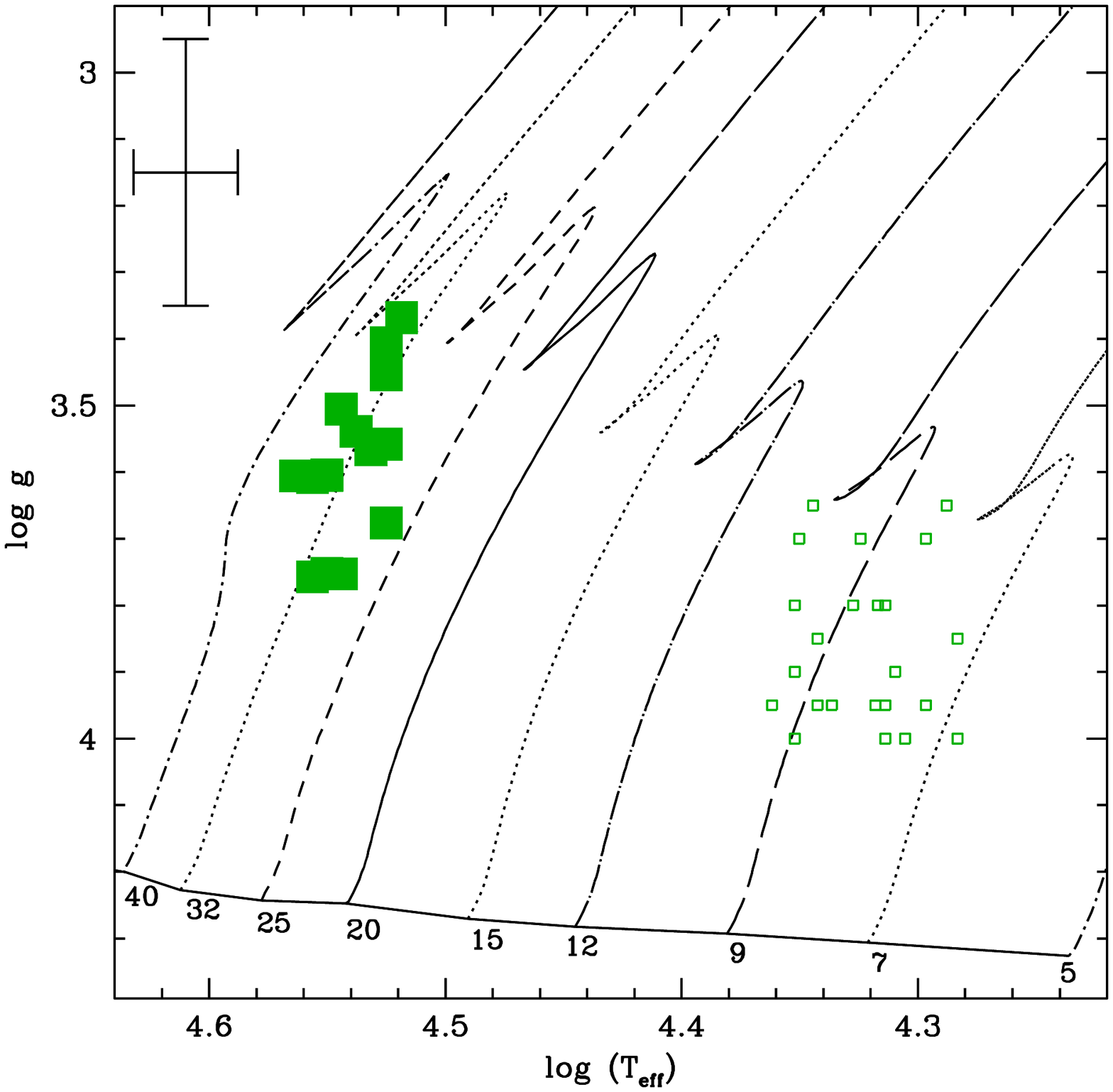}
\caption{As Fig.\ \ref{fig_hr} with a selection of B stars from \citet{hunter09} and \citet{np12} shown by open squares.}
\label{hr_B}
\end{figure}

The initial goal of the present study was to study the surface abundances of a homogeneous sample of O stars to test the effects of rotation. In particular, we wanted to restrict the effects of initial mass, metallicity and age. Indeed, among other effects, surface chemical enrichment is predicted to increase with initial mass \citep[e.g.,][]{brott11a}. \citet{mimesO} showed that among O stars this seems to be confirmed by observations. We can use our results to further test this mass dependence. For that, we have selected Galactic B stars from the studies of \citet{hunter09} and \citet{np12}. We have chosen objects located on the second half of the main sequence (3.7 $<$ \logg\ $<$ 4.0) and in a relatively narrow temperature range (20000 $<$ \teff\ $<$ 25000 K). These objects are placed in the \logg\ -- \teff\ diagram in Fig.\ \ref{hr_B}. They correspond to stars with initial masses between 7 and 10 \msun, a mass range about four times smaller than that of our O7-8 giant stars. 

\begin{figure}[t]
\centering
\includegraphics[width=9cm]{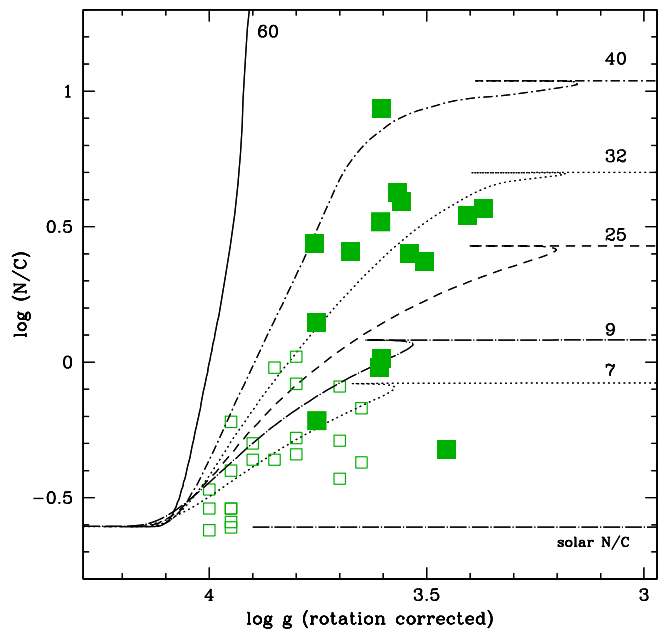}
\caption{As Fig.\ \ref{NC_logg} with a selection of B stars from \citet{hunter09} and \citet{np12} shown by open squares. Error bars are not indicated for clarity.}
\label{nc_logg_B}
\end{figure}

Figure\ \ref{nc_logg_B} presents the log(N/C) -- \logg\ diagram for both samples. There is a clear difference between O and B stars. O stars have log(N/C)$=0.33\pm0.34$ while B stars have log(N/C)$=-0.36\pm0.21$. Even though there is a range of initial rotational velocities in both samples, one can say that on average, the B-stars sample display N/C ratios almost five times smaller than the O-stars sample. According to Fig.\ \ref{nc_logg_B} this is also predicted by the models of \citet{ek12}. We thus conclude that the dependence of surface chemical enrichment on initial mass predicted by stellar evolution with rotation is verified observationally.

\section{Conclusion}
\label{s_conc}

We have performed a spectroscopic study of fifteen Galactic O7-8 giant stars. They were selected to represent a homogeneous sample in terms of initial mass, evolutionary state and metallicity. Our goal was to test the effect of rotational velocity on the surface chemical composition, minimizing the other effects due to mass, age and metallicity. Optical spectra have been collected from the IACOB and OWN spectroscopic surveys. We selected giant stars because they can still show a wide range of rotational velocities (unlike post-main sequence stars) and at the same time they are evolved enough to show surface chemical enrichment. The spectroscopic analysis was conducted with atmosphere models computed with the code CMFGEN. The main results are as follows.

All stars are located on the second half of the main sequence and have an initial mass between 25 and 40 \msun\ \citep[according to the tracks of][]{ek12}. The sample is thus relatively homogeneous. The distribution of projected rotational velocities of the sample covers the range 50-300 \kms. 

Most stars with \vsini\ larger than 100 \kms\ show a significant enrichment, with N/C values about ten times larger than the initial, solar ratio. Stars with lower projected rotational velocities show a wide range of N/C, from about two up to ten times the initial ratio. These results indicate that stars with low \vsini\ can be chemically evolved. This is similar to what was found by \citet{hunter08,hunter09} and \citet{rg12} for Magellanic Clouds OB main sequence stars. 

The evolutionary models of \citet{ek12} reproduce very well the position of the O7-8 giant stars in the log(N/C) -- \logg\ diagram. The main results of our study is that they also account for these stars in the log(N/C) -- \vsini\ diagram. Equatorial velocities in models with masses larger than 30 \msun\ decrease sufficiently during the main sequence to reach the range of observed low \vsini. At the same time, surface chemical enrichment is strong enough so that slowly rotating stars with high N/C ratios are reproduced. This is attributed to the combination of differential rotation and wind braking. Models computed by \citet{cl13} use a different diffusion coefficient for shear mixing compared to \citet{ek12} and consequently remain with too high velocities to provide a consistent explanation of all properties of our sample stars. 

The surface chemical composition of our sample of O7-8 giants shows N/C ratios larger by a factor five compared to a sample of Galactic B stars from \citet{hunter09} and \citet{np12}. The latter sample consists of stars with initial masses about four times smaller than the O7-8 giants, but is similarly made of stars located on the second half of the main sequence. Consequently, a clear trend of higher surface chemical enrichment at higher masses is established.

No clear correlation between surface chemical composition and projected rotational velocity was obtained.
The present study, however, indicates that this is not inconsistent with stellar evolution in the mass range we probed (25-40 \msun). Evolutionary models with an advecto-diffusive treatment of rotation are able to provide a consistent picture of stellar evolution on the main sequence at solar metallicity when they use the prescription by \citet{maeder97} for the shear mixing. Our results are based on a limited number of stars. Future investigations of larger samples should clarify whether entire stellar populations can be accounted for by such models.

\section{Acknowledgments}

We thank John Hillier for making CMFGEN available to the community. We acknowledge interesting discussions with Cyril Georgy. SS-D acknowledges funding by the Spanish Ministry of Economy and Competitiveness (MINECO) under the grants AYA2010-21697-C05-01, AYA2012-39364-C02-01, and Severo Ochoa SEV-2011-0187, and by the Canary Islands Government under grant PID2010119.  RB acknowledges support from FONDECYT Regular Project 1140076. RG was supported by grant PIP 112-201201-00298 (CONICET). We thank the referee, Ines Brott, for a timely report.

\bibliographystyle{aa}
\bibliography{rotOIII}

\begin{thebibliography}{39}
\expandafter\ifx\csname natexlab\endcsname\relax\def\natexlab#1{#1}\fi

\bibitem[{{Barb{\'a}} {et~al.}(2010){Barb{\'a}}, {Gamen}, {Arias}, {Morrell},
  {Ma{\'{\i}}z Apell{\'a}niz}, {Alfaro}, {Walborn}, \& {Sota}}]{barba10}
{Barb{\'a}}, R.~H., {Gamen}, R., {Arias}, J.~I., {et~al.} 2010, in Revista
  Mexicana de Astronomia y Astrofisica, vol.~27, Vol.~38, Revista Mexicana de
  Astronomia y Astrofisica Conference Series, 30--32

\bibitem[{{Bouret} {et~al.}(2012){Bouret}, {Hillier}, {Lanz}, \&
  {Fullerton}}]{jc12}
{Bouret}, J.-C., {Hillier}, D.~J., {Lanz}, T., \& {Fullerton}, A.~W. 2012,
  \aap, 544, A67

\bibitem[{{Bouret} {et~al.}(2013){Bouret}, {Lanz}, {Martins}, {Marcolino},
  {Hillier}, {Depagne}, \& {Hubeny}}]{jc13}
{Bouret}, J.-C., {Lanz}, T., {Martins}, F., {et~al.} 2013, \aap, 555, A1

\bibitem[{{Brott} {et~al.}(2011{\natexlab{a}}){Brott}, {de Mink}, {Cantiello},
  {Langer}, {de Koter}, {Evans}, {Hunter}, {Trundle}, \& {Vink}}]{brott11a}
{Brott}, I., {de Mink}, S.~E., {Cantiello}, M., {et~al.} 2011{\natexlab{a}},
  \aap, 530, A115

\bibitem[{{Brott} {et~al.}(2011{\natexlab{b}}){Brott}, {Evans}, {Hunter}, {de
  Koter}, {Langer}, {Dufton}, {Cantiello}, {Trundle}, {Lennon}, {de Mink},
  {Yoon}, \& {Anders}}]{brott11b}
{Brott}, I., {Evans}, C.~J., {Hunter}, I., {et~al.} 2011{\natexlab{b}}, \aap,
  530, A116

\bibitem[{{Chieffi} \& {Limongi}(2013)}]{cl13}
{Chieffi}, A. \& {Limongi}, M. 2013, \apj, 764, 21

\bibitem[{{Chiosi} \& {Maeder}(1986)}]{cm86}
{Chiosi}, C. \& {Maeder}, A. 1986, \araa, 24, 329

\bibitem[{{de Mink} {et~al.}(2014){de Mink}, {Sana}, {Langer}, {Izzard}, \&
  {Schneider}}]{demink14}
{de Mink}, S.~E., {Sana}, H., {Langer}, N., {Izzard}, R.~G., \& {Schneider},
  F.~R.~N. 2014, \apj, 782, 7

\bibitem[{{Ekstr{\"o}m} {et~al.}(2012){Ekstr{\"o}m}, {Georgy}, {Eggenberger},
  {Meynet}, {Mowlavi}, {Wyttenbach}, {Granada}, {Decressin}, {Hirschi},
  {Frischknecht}, {Charbonnel}, \& {Maeder}}]{ek12}
{Ekstr{\"o}m}, S., {Georgy}, C., {Eggenberger}, P., {et~al.} 2012, \aap, 537,
  A146

\bibitem[{{Endal} \& {Sofia}(1978)}]{es78}
{Endal}, A.~S. \& {Sofia}, S. 1978, \apj, 220, 279

\bibitem[{{Heger} {et~al.}(2000){Heger}, {Langer}, \& {Woosley}}]{hlw00}
{Heger}, A., {Langer}, N., \& {Woosley}, S.~E. 2000, \apj, 528, 368

\bibitem[{{Hillier} {et~al.}(2003){Hillier}, {Lanz}, {Heap}, {Hubeny}, {Smith},
  {Evans}, {Lennon}, \& {Bouret}}]{hil03}
{Hillier}, D.~J., {Lanz}, T., {Heap}, S.~R., {et~al.} 2003, \apj, 588, 1039

\bibitem[{{Hillier} \& {Miller}(1998)}]{hm98}
{Hillier}, D.~J. \& {Miller}, D.~L. 1998, \apj, 496, 407

\bibitem[{{Hunter} {et~al.}(2009){Hunter}, {Brott}, {Langer}, {Lennon},
  {Dufton}, {Howarth}, {Ryans}, {Trundle}, {Evans}, {de Koter}, \&
  {Smartt}}]{hunter09}
{Hunter}, I., {Brott}, I., {Langer}, N., {et~al.} 2009, \aap, 496, 841

\bibitem[{{Hunter} {et~al.}(2008){Hunter}, {Brott}, {Lennon}, {Langer},
  {Dufton}, {Trundle}, {Smartt}, {de Koter}, {Evans}, \& {Ryans}}]{hunter08}
{Hunter}, I., {Brott}, I., {Lennon}, D.~J., {et~al.} 2008, \apjl, 676, L29

\bibitem[{{Langer}(2012)}]{langer12}
{Langer}, N. 2012, \araa, 50, 107

\bibitem[{{Maeder}(1997)}]{maeder97}
{Maeder}, A. 1997, \aap, 321, 134

\bibitem[{{Maeder} \& {Meynet}(2000)}]{mm00}
{Maeder}, A. \& {Meynet}, G. 2000, \araa, 38, 143

\bibitem[{{Maeder} {et~al.}(2009){Maeder}, {Meynet}, {Ekstr{\"o}m}, \&
  {Georgy}}]{maeder09}
{Maeder}, A., {Meynet}, G., {Ekstr{\"o}m}, S., \& {Georgy}, C. 2009,
  Communications in Asteroseismology, 158, 72

\bibitem[{{Maeder} {et~al.}(2014){Maeder}, {Przybilla}, {Nieva}, {Georgy},
  {Meynet}, {Ekstr{\"o}m}, \& {Eggenberger}}]{maeder14}
{Maeder}, A., {Przybilla}, N., {Nieva}, M.-F., {et~al.} 2014, \aap, 565, A39

\bibitem[{{Maeder} \& {Zahn}(1998)}]{mz98}
{Maeder}, A. \& {Zahn}, J.-P. 1998, \aap, 334, 1000

\bibitem[{{Martins} {et~al.}(2015{\natexlab{a}}){Martins}, {Herv{\'e}},
  {Bouret}, {Marcolino}, {Wade}, {Neiner}, {Alecian}, {Grunhut}, \&
  {Petit}}]{mimesO}
{Martins}, F., {Herv{\'e}}, A., {Bouret}, J.-C., {et~al.} 2015{\natexlab{a}},
  \aap, 575, A34

\bibitem[{{Martins} {et~al.}(2012){Martins}, {Mahy}, {Hillier}, \&
  {Rauw}}]{ngc2244}
{Martins}, F., {Mahy}, L., {Hillier}, D.~J., \& {Rauw}, G. 2012, \aap, 538, A39

\bibitem[{{Martins} {et~al.}(2015{\natexlab{b}}){Martins}, {Marcolino},
  {Hillier}, {Donati}, \& {Bouret}}]{varnarval}
{Martins}, F., {Marcolino}, W., {Hillier}, D.~J., {Donati}, J.-F., \& {Bouret},
  J.-C. 2015{\natexlab{b}}, \aap, 574, A142

\bibitem[{{Martins} \& {Palacios}(2013)}]{mp13}
{Martins}, F. \& {Palacios}, A. 2013, \aap, 560, A16

\bibitem[{{Martins} {et~al.}(2005){Martins}, {Schaerer}, \& {Hillier}}]{msh05}
{Martins}, F., {Schaerer}, D., \& {Hillier}, D.~J. 2005, \aap, 436, 1049

\bibitem[{{Meynet} \& {Maeder}(2000)}]{mema00}
{Meynet}, G. \& {Maeder}, A. 2000, \aap, 361, 101

\bibitem[{{Morel} {et~al.}(2006){Morel}, {Butler}, {Aerts}, {Neiner}, \&
  {Briquet}}]{morel06}
{Morel}, T., {Butler}, K., {Aerts}, C., {Neiner}, C., \& {Briquet}, M. 2006,
  \aap, 457, 651

\bibitem[{{Morel} {et~al.}(2008){Morel}, {Hubrig}, \& {Briquet}}]{morel08}
{Morel}, T., {Hubrig}, S., \& {Briquet}, M. 2008, \aap, 481, 453

\bibitem[{{Nieva} \& {Przybilla}(2012)}]{np12}
{Nieva}, M.-F. \& {Przybilla}, N. 2012, \aap, 539, A143

\bibitem[{{Rivero Gonz{\'a}lez} {et~al.}(2012){Rivero Gonz{\'a}lez}, {Puls},
  {Najarro}, \& {Brott}}]{rg12}
{Rivero Gonz{\'a}lez}, J.~G., {Puls}, J., {Najarro}, F., \& {Brott}, I. 2012,
  \aap, 537, A79

\bibitem[{{Sim{\'o}n-D{\'{\i}}az} \& {Herrero}(2014)}]{sergio14}
{Sim{\'o}n-D{\'{\i}}az}, S. \& {Herrero}, A. 2014, \aap, 562, A135

\bibitem[{{Sim{\'o}n-D{\'{\i}}az} {et~al.}(2015){Sim{\'o}n-D{\'{\i}}az},
  {Negueruela}, {Ma{\'{\i}}z Apell{\'a}niz}, {Castro}, {Herrero}, {Garcia},
  {P{\'e}rez-Prieto}, {Caon}, {Alacid}, {Camacho}, {Dorda}, {Godart},
  {Gonz{\'a}lez-Fern{\'a}ndez}, {Holgado}, \& {R{\"u}bke}}]{ss15}
{Sim{\'o}n-D{\'{\i}}az}, S., {Negueruela}, I., {Ma{\'{\i}}z Apell{\'a}niz}, J.,
  {et~al.} 2015, in Highlights of Spanish Astrophysics VIII, ed. A.~J.
  {Cenarro}, F.~{Figueras}, C.~{Hern{\'a}ndez-Monteagudo}, J.~{Trujillo Bueno},
  \& L.~{Valdivielso}, 576--581

\bibitem[{{Sota} {et~al.}(2014){Sota}, {Ma{\'{\i}}z Apell{\'a}niz}, {Morrell},
  {Barb{\'a}}, {Walborn}, {Gamen}, {Arias}, \& {Alfaro}}]{sota14}
{Sota}, A., {Ma{\'{\i}}z Apell{\'a}niz}, J., {Morrell}, N.~I., {et~al.} 2014,
  \apjs, 211, 10

\bibitem[{{Sota} {et~al.}(2011){Sota}, {Ma{\'{\i}}z Apell{\'a}niz}, {Walborn},
  {Alfaro}, {Barb{\'a}}, {Morrell}, {Gamen}, \& {Arias}}]{sota11}
{Sota}, A., {Ma{\'{\i}}z Apell{\'a}niz}, J., {Walborn}, N.~R., {et~al.} 2011,
  \apjs, 193, 24

\bibitem[{{Talon} \& {Zahn}(1997)}]{tz97}
{Talon}, S. \& {Zahn}, J.-P. 1997, \aap, 317, 749

\bibitem[{{Vink} {et~al.}(2001){Vink}, {de Koter}, \& {Lamers}}]{vink01}
{Vink}, J.~S., {de Koter}, A., \& {Lamers}, H.~J.~G.~L.~M. 2001, \aap, 369, 574

\bibitem[{{Walborn} {et~al.}(2011){Walborn}, {Ma{\'{\i}}z Apell{\'a}niz},
  {Sota}, {Alfaro}, {Morrell}, {Barb{\'a}}, {Arias}, \& {Gamen}}]{walborn11}
{Walborn}, N.~R., {Ma{\'{\i}}z Apell{\'a}niz}, J., {Sota}, A., {et~al.} 2011,
  \aj, 142, 150

\bibitem[{{Zahn}(1992)}]{zahn92}
{Zahn}, J.-P. 1992, \aap, 265, 115

\end{thebibliography}

\newpage

\begin{appendix}

\section{Spectral variability}
\label{ap_var}

\begin{figure*}[]
\centering
\includegraphics[width=16cm]{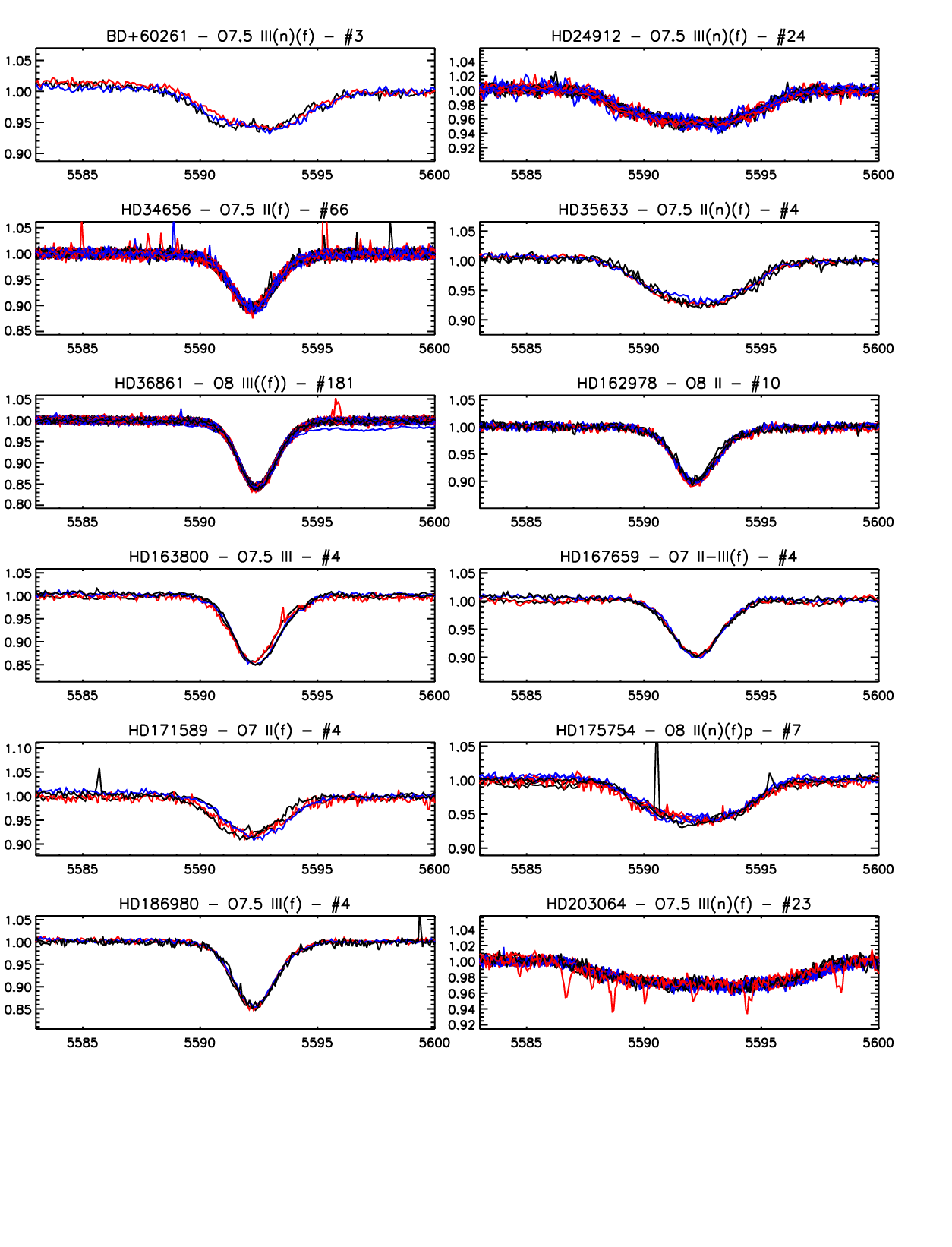}
\caption{Temporal variability of \ion{O}{iii}~5592 for stars having more than one spectrum available. The number of spectra is indicated for each star. Different colors are used for different dates of observation.}
\label{fig_var}
\end{figure*}

In Fig.\ \ref{fig_var} we show the \ion{O}{iii}~5592 line for stars for which several spectra are available. This allows us to check if variabiity is observed, and to exclude obvious binary systems. \ion{O}{iii}~5592 is mainly a photospheric line \citep[e.g.,][]{varnarval}, ensuring that wind variability is limited and that spectral variations are due to the star's motion or to photospheric processes. 

\pagebreak

\section{Best fits to the observed spectra}
\label{ap_fit}

In this section we show the best fits to the observed spectra af all target stars. The CMFGEN models are shown in red and the observed data in black. The fits are of good quality. Small discrepancies exist and result mainly from the adoption of stellar parameters and abundances aimed at representing an entire set of lines, not individual lines. For instance, \ion{O}{iii}~5592 is sometimes too strong or too weak in our models, due to the adoption of an oxygen abundance best accounting for both \ion{O}{iii}~3962 and \ion{O}{iii}~5592, and sometimes \ion{O}{iii}~3791 too (see Sect.\ \ref{s_mod}). 

\begin{figure}[]
\centering
\includegraphics[width=9cm]{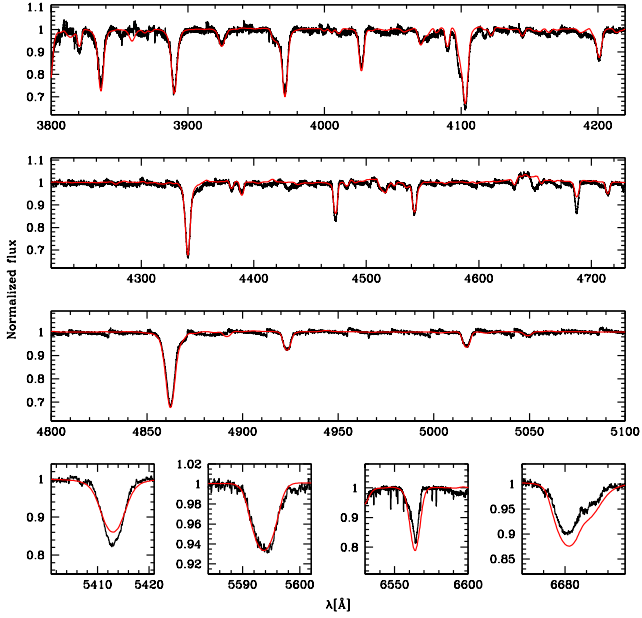}
\caption{Best fit (red) of the observed spectrum (black) of BD~+60261.}
\label{fit_bd60261}
\end{figure}

\begin{figure}[]
\centering
\includegraphics[width=9cm]{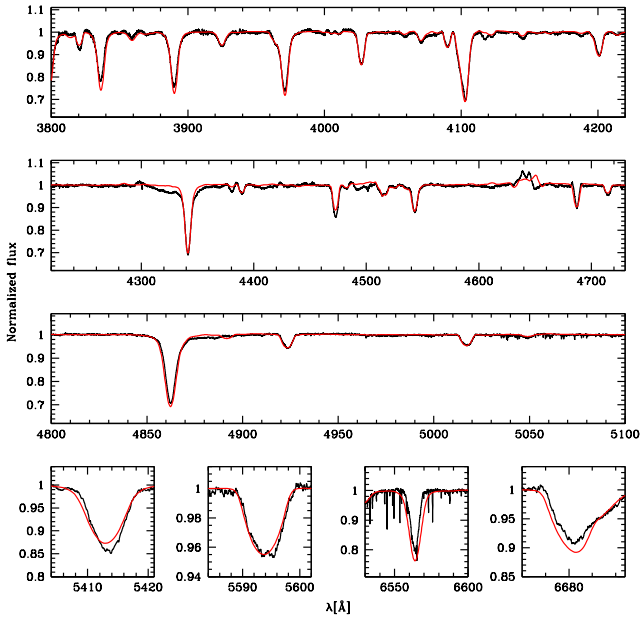}
\caption{Best fit (red) of the observed spectrum (black) of HD~24912.}
\label{fit_24912}
\end{figure}

\begin{figure}[]
\centering
\includegraphics[width=9cm]{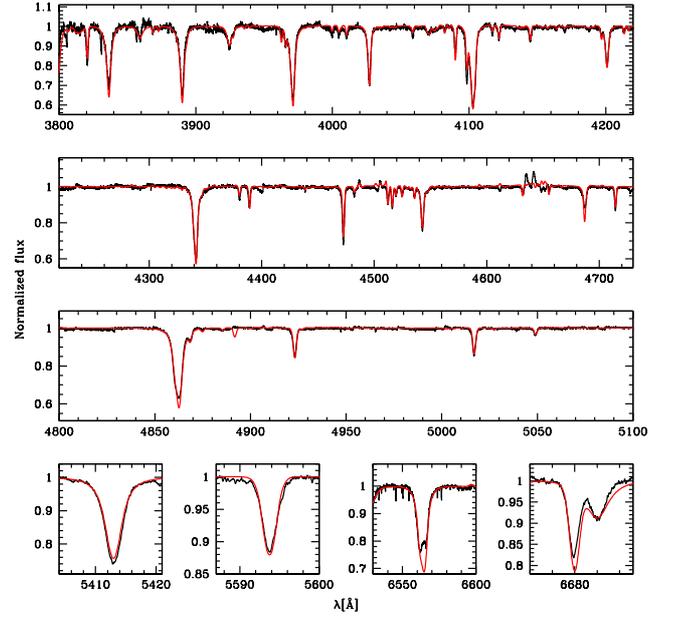}
\caption{Best fit (red) of the observed spectrum (black) of HD~34656.}
\label{fit_34656}
\end{figure}

\begin{figure}[]
\centering
\includegraphics[width=9cm]{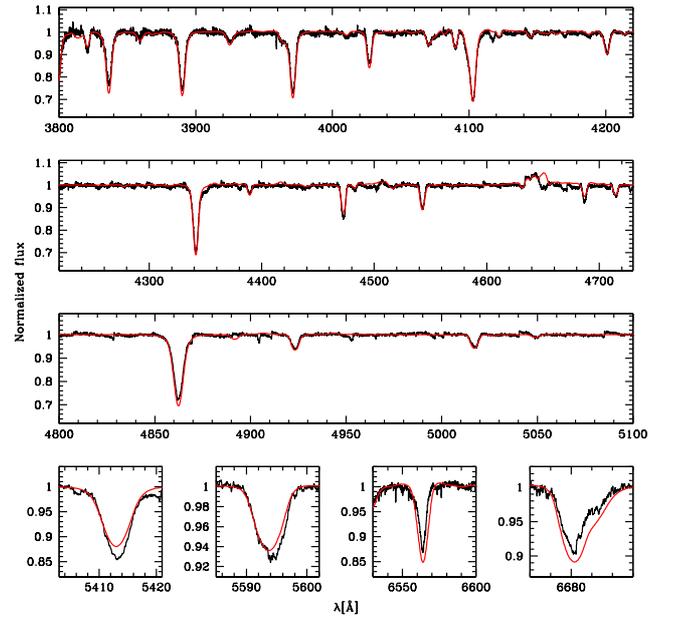}
\caption{Best fit (red) of the observed spectrum (black) of HD~35633.}
\label{fit_35633}
\end{figure}

\begin{figure}[]
\centering
\includegraphics[width=9cm]{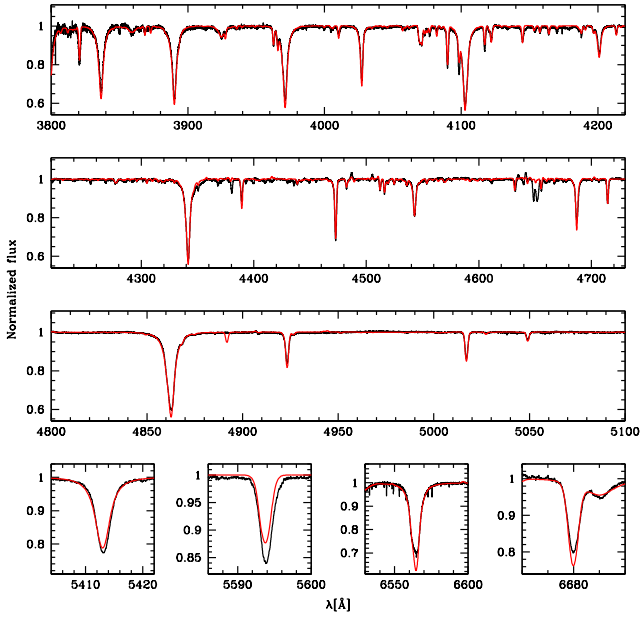}
\caption{Best fit (red) of the observed spectrum (black) of HD~36861.}
\label{fit_36861}
\end{figure}

\begin{figure}[]
\centering
\includegraphics[width=9cm]{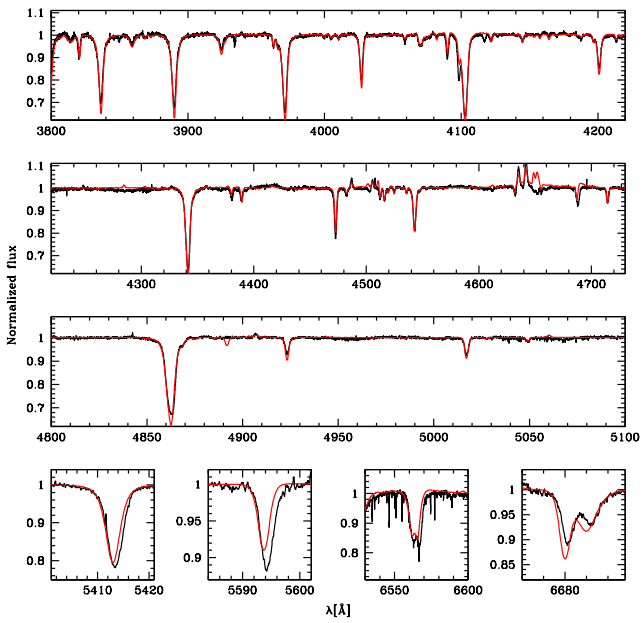}
\caption{Best fit (red) of the observed spectrum (black) of HD~94963.}
\label{fit_94963}
\end{figure}

\begin{figure}[]
\centering
\includegraphics[width=9cm]{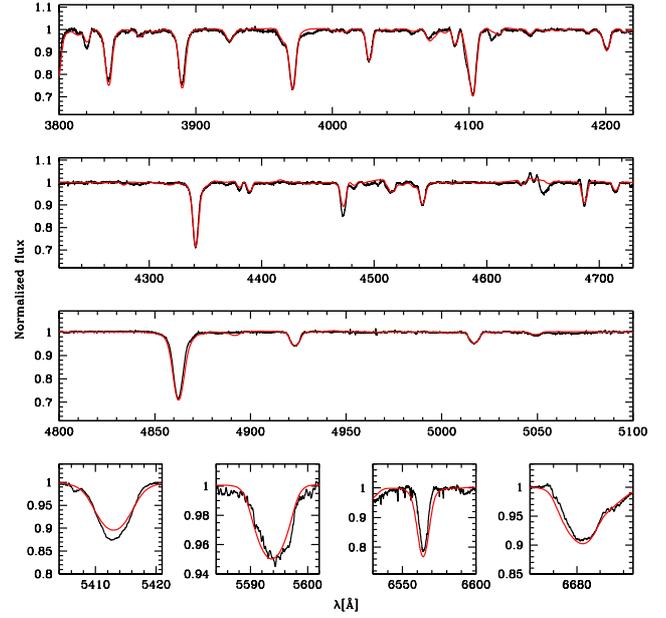}
\caption{Best fit (red) of the observed spectrum (black) of HD~97434.}
\label{fit_97434}
\end{figure}

\begin{figure}[]
\centering
\includegraphics[width=9cm]{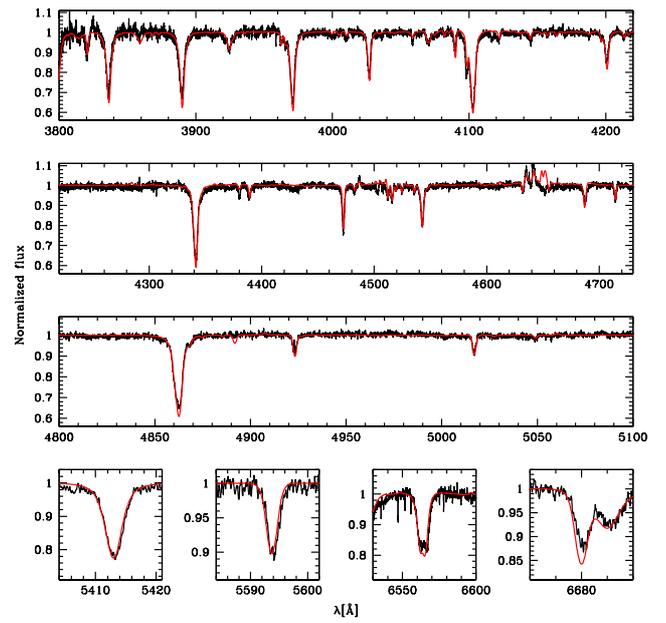}
\caption{Best fit (red) of the observed spectrum (black) of HD~151515.}
\label{fit_151515}
\end{figure}

\begin{figure}[]
\centering
\includegraphics[width=9cm]{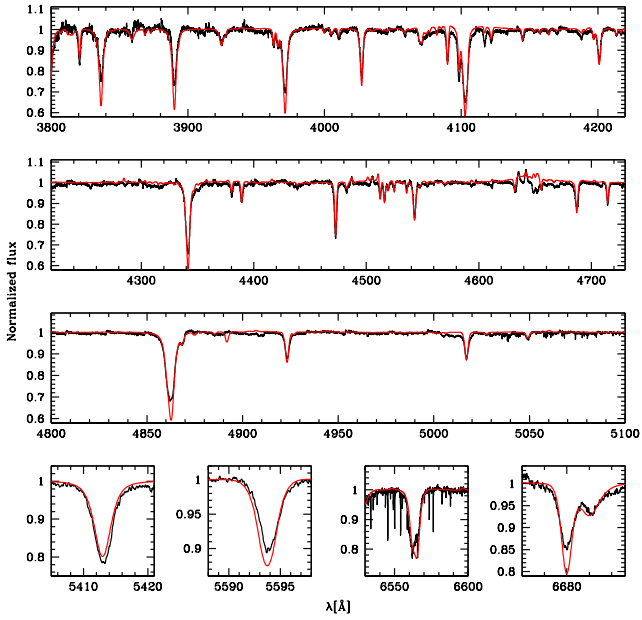}
\caption{Best fit (red) of the observed spectrum (black) of HD~162978.}
\label{fit_162978}
\end{figure}

\begin{figure}[]
\centering
\includegraphics[width=9cm]{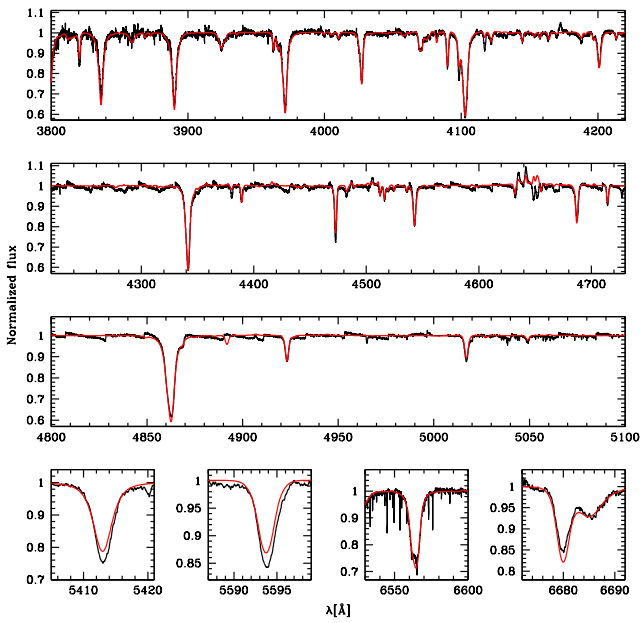}
\caption{Best fit (red) of the observed spectrum (black) of HD~163800.}
\label{fit_163800}
\end{figure}

\begin{figure}[]
\centering
\includegraphics[width=9cm]{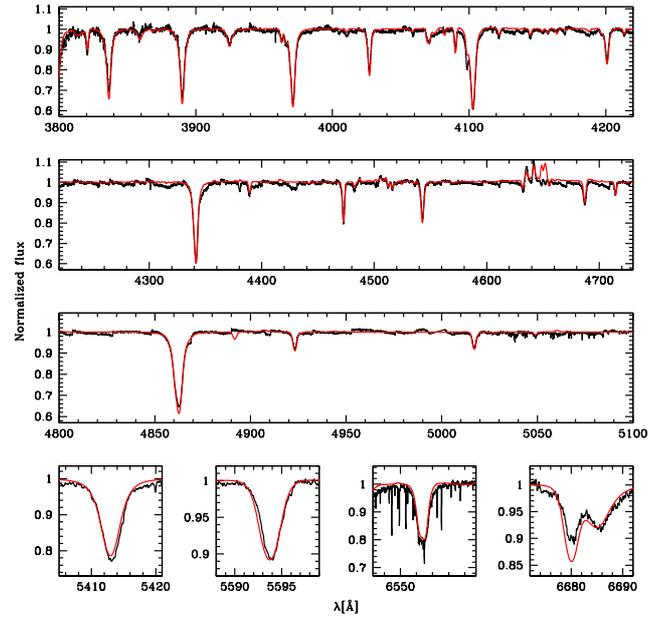}
\caption{Best fit (red) of the observed spectrum (black) of HD~167659.}
\label{fit_167659}
\end{figure}

\begin{figure}[]
\centering
\includegraphics[width=9cm]{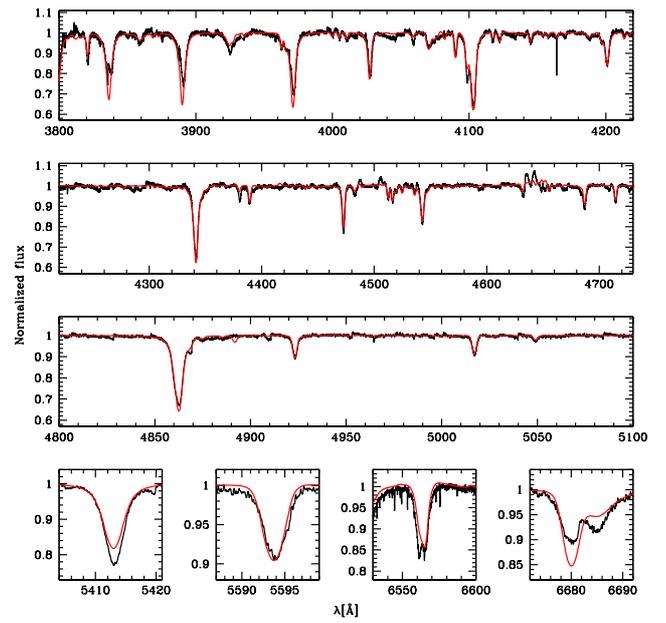}
\caption{Best fit (red) of the observed spectrum (black) of HD~171589.}
\label{fit_171589}
\end{figure}

\begin{figure}[]
\centering
\includegraphics[width=9cm]{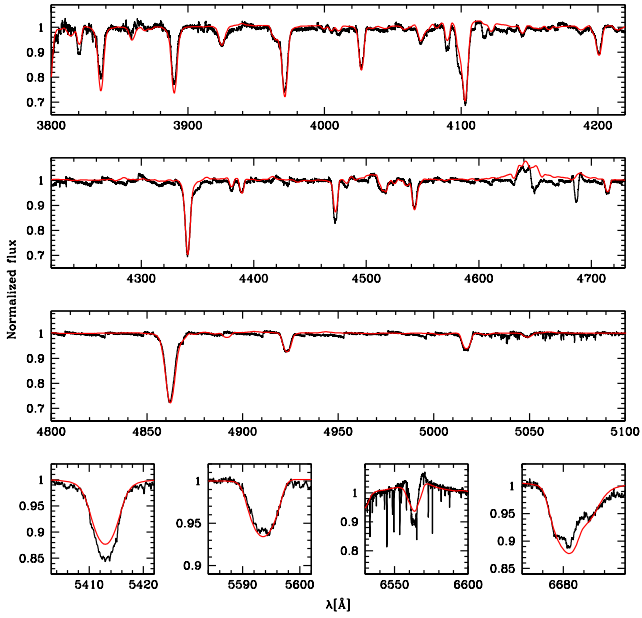}
\caption{Best fit (red) of the observed spectrum (black) of HD~175754.}
\label{fit_175754}
\end{figure}

\begin{figure}[]
\centering
\includegraphics[width=9cm]{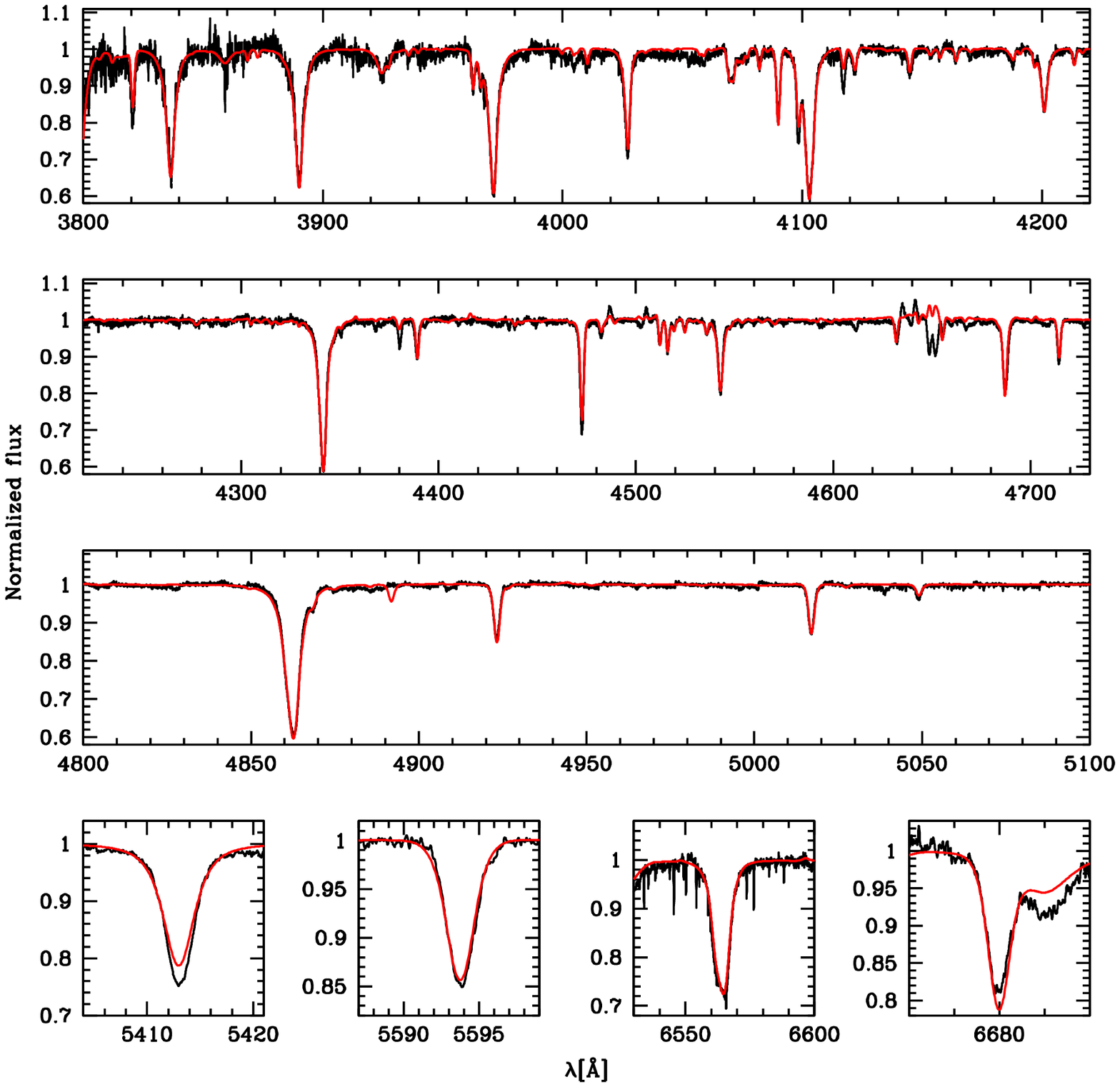}
\caption{Best fit (red) of the observed spectrum (black) of HD~186980.}
\label{fit_186980}
\end{figure}

\begin{figure}[]
\centering
\includegraphics[width=9cm]{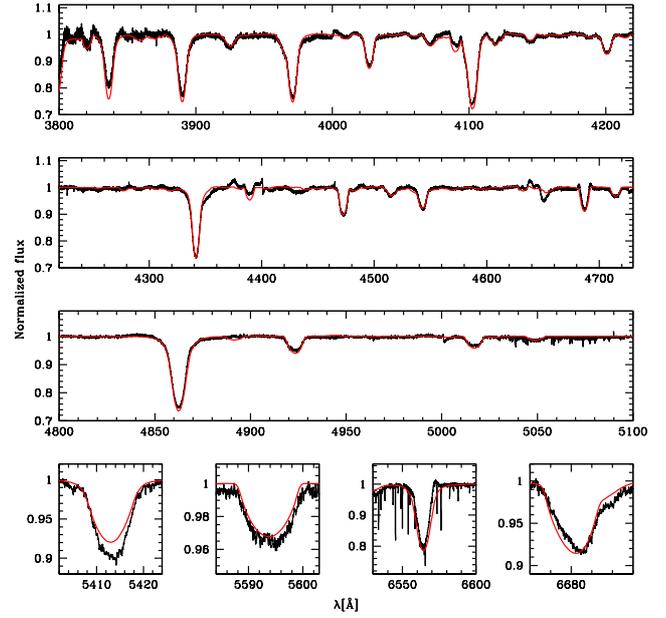}
\caption{Best fit (red) of the observed spectrum (black) of HD~203064.}
\label{fit_203064}
\end{figure}

\end{appendix}

\end{document}